\documentclass[11pt, letterpaper]{article}
\usepackage{graphicx}
\usepackage[square,comma,numbers]{natbib}
\usepackage{amsmath}
\usepackage{amssymb}
\usepackage{amsmath}
\usepackage{amsfonts}
\usepackage{mathtools}
\usepackage[super]{nth}
\usepackage{setspace}
\usepackage{enumitem} 
\usepackage{epstopdf}
\usepackage{color}
\usepackage[letterpaper,left=1in,right=1in,top=1in,bottom=1in]{geometry}
\usepackage[ruled,vlined]{algorithm2e}
\usepackage{multirow}
\usepackage{longtable}
\usepackage{rotating}
\usepackage{mathrsfs}
\usepackage{subcaption}
\usepackage{float}
\usepackage{graphicx}
\usepackage{booktabs}
\usepackage{caption}
\usepackage{bm}
\usepackage{algorithmic}
\usepackage{tikz}
\usepackage{makecell}
\usetikzlibrary{shapes,arrows,positioning,calc}

\title{\Large \bf Integrating Machine Learning Paradigms and Mixed-Integer Model Predictive Control for Irrigation Scheduling}
\author{
    \centerline{\normalsize Bernard T. Agyeman$^{a}$, Mohamed Naouri$^{b}$, Willemijn Appels$^{b}$, Jinfeng Liu$^{a}$\thanks{Corresponding author: J. Liu. Tel: +1-780-492-1317. Fax: +1-780-492-2881. Email: jinfeng@ualberta.ca.}, Sirish L. Shah$^{a}$}
\vspace{5mm}\\
\centerline{\small $^{a}$Department of Chemical \& Materials Engineering, University of Alberta,}\\
\centerline{\small Edmonton, AB, Canada T6G 1H9.}\\
\centerline{\small $^{b}$Centre for Applied Research, Innovation, and Entrepreneurship, Lethbridge College,}\\
\centerline{\small Lethbridge, AB, Canada T1K 1L6.}
}
\allowdisplaybreaks

\begin{document}
\date{}
\maketitle
%\doublespacing

%\onehalfspacing
\setstretch{1.5}
{}
\begin{abstract}
The agricultural sector currently faces significant challenges in water resource conservation and crop yield optimization, primarily due to concerns over freshwater scarcity. Traditional irrigation scheduling methods often prove inadequate in meeting the needs of large-scale irrigation systems. To address this issue, this paper proposes a predictive irrigation scheduler that leverages the three paradigms of machine learning to optimize irrigation schedules. The proposed scheduler employs the k-means clustering approach to divide the field into distinct irrigation management zones based on soil hydraulic parameters and topology information. Furthermore, a long short-term memory network is employed to develop dynamic models for each management zone, enabling accurate predictions of soil moisture dynamics. Formulated as a mixed-integer model predictive control problem, the scheduler aims to maximize water uptake while minimizing overall water consumption and irrigation costs. To tackle the mixed-integer optimization challenge, the proximal policy optimization algorithm is utilized to train a reinforcement learning agent responsible for making daily irrigation decisions. To evaluate the performance of the proposed scheduler, a 26.4-hectare field in Lethbridge, Canada, was chosen as a case study for the 2015 and 2022 growing seasons. The results demonstrate the superiority of the proposed scheduler compared to a traditional irrigation scheduling method in terms of water use efficiency and crop yield improvement for both growing seasons. Notably, the proposed scheduler achieved water savings ranging from 6.4\% to 22.8\%, along with yield increases ranging from 2.3\% to 4.3\%.
\end{abstract}
\clearpage
\section{Introduction}
According to a report by the United Nations, agriculture accounts for approximately 70\% of the world's freshwater withdrawals, with the majority of the water being used for irrigation activities~\cite{unwater2015}. However, with the global freshwater shortage crisis worsening due to factors such as rapid population growth and climate change, there is a pressing need for effective water management strategies, particularly in relation to irrigation activities, in order to address the growing freshwater shortages.

While irrigation is essential for optimal plant development in areas with inadequate rainfall, it is equally important to irrigate plants with the right amount of water in a timely manner. This necessitates the implementation of effective irrigation scheduling operations on a daily basis, with a planning horizon spanning a few days or weeks~\cite{gu2020irrigation}. In conventional irrigation management, scheduling operations are typically implement in an open-loop fashion, which means that there is no direct connection between the supplied irrigation rate and the soil water status. However, open-loop systems tend to be imprecise and do not guarantee optimal plant yield and enhanced water use efficiency. To address these limitations, closed-loop irrigation methods have been recommended. By closing the irrigation decision-making loop, it is possible to overcome the drawbacks of open-loop irrigation systems and improve plant yield and water use efficiency. Machine learning techniques and model predictive control have emerged as valuable tools in designing closed-loop irrigation schedulers.

To optimize water-use efficiency and crop productivity, it is crucial to consider the inherent variability in fields resulting from physical, chemical, and biological processes that occur in soil. Implementing a variable rate irrigation scheduling approach is recommended, particularly for fields with significant variability. This involves dividing the field into distinct irrigation management zones that share similar soil properties and cropping conditions, enabling a targeted and customized irrigation management strategy. Unsupervised machine learning techniques, such as the k-means method and fuzzy k-means method, have been utilized to delineate these management zones by utilizing various attributes like topography, soil texture, remote sensing data, yield maps, and forest cover images~\cite{haghverdi2015perspectives,stafford1999using,xin2009determination,li2007delineation}. By integrating multiple data sources and leveraging clustering algorithms, accurate and reliable delineations of the management zones can be generated~\cite{cordoba2013subfield}.

Mechanistic agro-hydrological models, such as the Richards equation and AquaCrop model, are widely used to enhance the reliability and precision of closed-loop irrigation schedulers~\cite{park2009receding, delgoda2016irrigation}. However, these models have notable drawbacks. Firstly, they rely on time-consuming calibration processes that demand significant effort. Moreover, their computational complexity makes them impractical, particularly when prompt real-time irrigation decisions are required. To address these challenges, data-driven models that employ supervised machine learning techniques offer a promising solution. These models take a different approach to irrigation scheduling by directly learning patterns and relationships from observed data. Among the data-driven models, feed-forward neural networks~\cite{gu2021neural} and long short-term memory networks (LSTMs)~\cite{adeyemi2018dynamic,agyeman2023lstm} have gained significant attention. These data-driven models have demonstrated their capability to enhance the efficiency and accuracy of closed-loop irrigation schedulers. Leveraging the power of machine learning algorithms, they efficiently learn from historical data and can thus be used to make real-time irrigation decisions.

Uncertainty in weather conditions presents a significant challenge for closed-loop irrigation scheduling. This challenge arises because weather fluctuations directly affect the water requirements of plants, leading to potential issues of over-watering or under-watering when weather predictions are inaccurate. To tackle this challenge, the Reinforcement Learning (RL) paradigm of machine learning has been used to develop robust closed-loop irrigation schedulers that can adapt effectively to uncertain weather conditions~\cite{chen2021reinforcement}.
Furthermore, accurate modeling is vital in the design of closed-loop irrigation systems, as the use of inaccurate models can result in suboptimal schedules and reduced crop yields. To address the issue of model inaccuracies, RL techniques have been employed. These techniques enable the irrigation system to learn from experience and adjust its schedules accordingly, enhancing its performance and optimizing water usage~\cite{ding2023optimizing}. Additionally, the design of closed-loop irrigation schedulers often involves manual interpretation of sensor data, which reduces the system's autonomy and requires human intervention. To overcome this limitation and achieve fully automatic closed-loop schedulers capable of achieving optimal or near-optimal control, RL techniques have been employed. By leveraging RL, these systems can learn and make decisions autonomously, minimizing the need for manual intervention and maximizing their efficiency~\cite{yang2020deep}.

The implementation of optimal control methods has significantly enhanced closed-loop irrigation scheduling by optimizing specific performance measures through the determination of water application depth and irrigation timing. Two mathematical techniques commonly used to address optimal control problems are Dynamic Programming (DP) and Model Predictive Control (MPC). DP has been previously employed in irrigation scheduling to generate irrigation schedules that maximize crop yield~\cite{naadimuthu1999heuristic}, but its practical application is limited due to the curse of dimensionality. In contrast, MPC, which is regarded as an effective implementation of the DP solution technique, has had a profound impact on control practice, particularly in managing constrained multivariable control problems. MPC synthesizes control inputs by repeatedly solving finite horizon optimal control problems across overlapping horizons. In the field of irrigation scheduling, MPC has been used to determine irrigation timing and the required irrigation volume for site-specific applications~\cite{delgoda2016irrigation,mccarthy2014simulation}. These studies aimed to maintain soil moisture in the root zone at predetermined set-points using MPC. However, the stability and feasibility of MPC tracking set-points are influenced by changes in the desired set-point, requiring MPC redesign with each set-point adjustment, which diminishes its computational viabilitye~\cite{bemporad1997nonlinear}. To ensure optimal plant growth, it is generally sufficient to maintain soil moisture within a specific range or zone, defined by the field capacity and a threshold value above the permanent wilting point of the soil. Consequently, an MPC approach with zone objectives, also known as zone control, becomes a natural choice for irrigation scheduling. In closed-loop irrigation scheduling, zone MPC has been utilized to schedule irrigation by minimizing the soil moisture deficit in the root zone and the overall irrigation volume~\cite{nahar2019closed}. Scheduling problems in irrigation inherently involve combinatorial challenges, as the allocation of limited resources to competing tasks over time necessitates discrete decisions. For example, in daily irrigation scheduling, the determination of the irrigation time reduces to a discrete decision of whether or not to perform the irrigation event each day within the planning horizon. To tackle these scheduling problems, our previous work proposed an LSTM-based mixed-integer MPC with zone control. This approach was developed to determine the daily irrigation decision and rate for both homogeneous and heterogeneous fields~\cite{agyeman2023lstm}.

This manuscript proposes a novel approach for irrigation scheduling, which combines all three machine learning paradigms with model predictive control. While previous studies have used only one or two of these techniques, we suggest a unified framework that capitalizes on the strengths of each method to achieve better accuracy and efficiency in irrigation scheduling. Our approach involves using unsupervised learning in a three-stage process to divide the irrigation field into distinct zones, followed by supervised learning to capture soil moisture dynamics in the resulting management zones. We then integrate these models into a mixed integer model predictive control framework with zone objectives, and employ reinforcement learning and the concept of a limiting irrigation management zone to eliminate the explicit determination of daily irrigation decisions. Specifically, we use the Proximal Policy Optimization (PPO) algorithm to develop a policy that provides the irrigation decision sequence over the planning horizon, resulting in a situation where only the daily irrigation rate is determined. 
% By combining these techniques, we aim to achieve more sustainable and cost-effective water use in agriculture.
The main contributions of this manuscript are as follows:
\begin{enumerate}
\item A three-stage process involving the k-means clustering approach to delineate the irrigation field into distinct management zone. 
% This process results in a representation this is consistent with the resolution of the irrigation implementing equipment 
\item The use of LSTM networks to identify multi-input-single-output data-driven models that capture the dynamics of soil moisture in the root-zone of the resulting irrigation management zones.
\item The use of the PPO algorithm and the concept of a limiting irrigation management zone to eliminate the explicit calculation of the daily discrete irrigation decision in the mixed-integer MPC.
\item A computationally efficient approach to calculate daily irrigation schedulers for fields that contain multiple irrigation management zones.
\item Evaluating the scheduler on a real field for two growing seasons and two management allowable depletions (MAD), where the scheduler outperforms a traditional irrigation scheduling approach in terms of water savings and predicted yield. 
\end{enumerate}

In another context, this manuscript can be seen as an expansion of our previous research, where we integrated the supervised learning paradigm of machine learning with a mixed-integer Model Predictive Control (MPC) to generate irrigation schedules for large-scale homogeneous and heterogeneous fields~\cite{agyeman2023lstm}. In our prior work, we introduced the concept of irrigation management zones, but we did not provide a systematic approach to delineate the field into these zones. Additionally, to improve the computational efficiency of the proposed scheduler, we approximated the binary variables arising in the mixed-integer MPC using a sigmoid function. However, this approximation introduced errors and occasionally led to an optimization problem that was poorly conditioned. In this study, we present a systematic method to delineate a field into irrigation management zones using the unsupervised learning paradigm. Furthermore, we propose the use of the Reinforcement Learning (RL) paradigm to obtain a policy that can be utilized to calculate the binary variables arising in the mixed-integer MPC. Moreover, in contrast to our previous work where we tested the proposed approach on simulated case studies, we evaluate the effectiveness of the proposed scheduler on a large-scale agricultural field in this manuscript. Additionally, we compare the schedules generated by our proposed approach with those obtained from a well-known scheduling method to assess its utility.

This paper expands on the preliminary results presented in~\cite{agyeman2023IFAC}. However, unlike~\cite{agyeman2023IFAC}, this study proposes an entirely new approach to identify the data-driven models for each irrigation management zone. In addition, this manuscript proposes an entirely new approach to evaluate the mixed-integer MPC. Finally, this manuscript evaluates the performance of the proposed scheduler on a large-scale agricultural field.

\section{Materials and methods}

\subsection{Unsupervised learning for management zone delineation} \label{delination}
\begin{figure}
\centering
\begin{tikzpicture}[node distance=1.5cm, scale=0.5]
% Stage 1: Crop Type-Based Management Zones
\node[draw,rectangle] (stage1) {Stage 1: Delineation According to Crop Type}; 
\node[draw,rectangle,below=of stage1] (stage2) {Stage 2: Delineation According to Elevation and Soil Hydraulic Parameters};
\node[draw,rectangle,below=of stage2] (stage3) {Stage 3: Mapping Homogeneous Areas to Irrigation Equipment Resolution};
% Arrows between stages
\draw[->] (stage1) -- (stage2);
\draw[->] (stage2) -- (stage3);
\end{tikzpicture}
\caption{A schematic diagram of the proposed management zone delineation process.}
\label{fig:delineation_processed}
\end{figure}
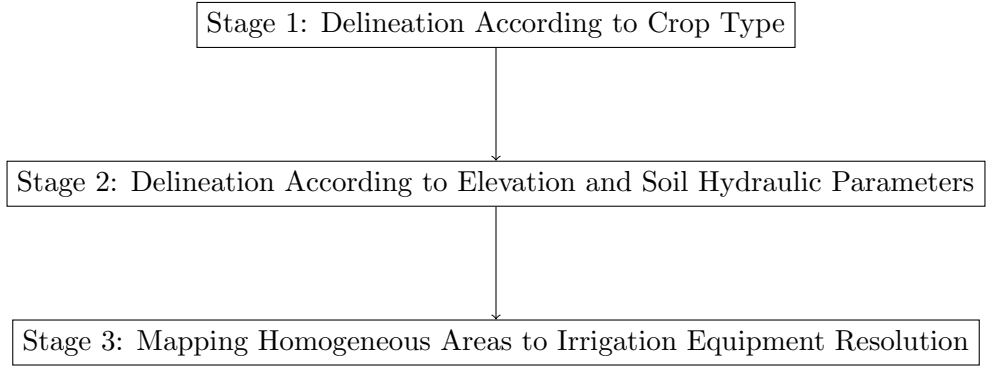

Spatial variability in irrigated fields is a common phenomenon due to the biological, chemical, and physical processes that occur in the soil. Efficient closed-loop irrigation systems require a thorough characterization of the spatial variability within a field, in order to maximize crop productivity and water use efficiency. Consequently, in this section, a three-stage process is proposed to delineate an irrigated field into distinct irrigation management zones. In the first stage of the process, the irrigated field is partitioned into management zones based on crop type. During the second stage, elevation information, soil hydraulic parameters, and the k-means clustering approach are utilized to further divide the resulting management zones into homogeneous areas. Finally, in the third stage, the homogeneous areas are mapped to the resolution of the field's irrigation implementing equipment. A schematic diagram of the proposed delineation process is shown  in Figure.~\ref{fig:delineation_processed}, and a detailed explanation of each stage of the process is provided below.

\begin{enumerate}
\item \textbf{Stage 1: delineation according to crop type.} Different crop types have varying water requirements, making crop water requirements a critical factor in irrigation scheduling. Since it is a common practice to find different crop types grown on irrigated fields, the initial delineation based on crop type provides a good basis for dividing the field into irrigation management zones. This division enables the three-stage process to account for crop-specific differences. 
% By creating management zones based on crop type, the irrigation management strategies can be tailored to meet the specific needs of each crop. 

\item \textbf{Stage 2: delineation according to elevation and soil-hydraulic parameters}. In this stage, additional information such as elevation and soil hydraulic parameters is used to further divide the management zones identified in Stage 1 into homogeneous areas. To ensure effective irrigation management, the management zone delineation process must account for elevation information, as it plays an important role in water movement within a field. Low-lying areas of a field are expected to accumulate excess water during heavy rains and irrigation, while high-elevation areas are prone to water loss from surface runoff, leading to reduced water penetration depths and plant growth restrictions. Additionally, soil hydraulic parameters also affect irrigation scheduling by governing soil water dynamics and plant available water. However, accessing these parameters at the field level remains a challenge. To deal with this challenge, auxiliary attributes such as soil electrical conductivity have been used to delineate irrigation management zones in the literature. This study utilizes a practical and computationally efficient technique proposed in our previous work~\cite{agyeman2022simultaneous,orouskhani2022simultaneous} to estimate soil moisture and hydraulic parameters simultaneously, which can provide the necessary hydraulic parameters for management zone delineation. In this approach, soil moisture and hydraulic parameter estimates are obtained by assimilating remotely sensed soil moisture observations into the Richards equation using the extended Kalman filtering technique, where the sensitivity analysis and orthogonal projection approaches are used to address issues of parameter estimability. The hydraulic parameters used in this work include saturated hydraulic conductivity, saturated moisture content, residual moisture content, and curve fitting parameters. The saturated hydraulic conductivity $\text{K}_{\text{s}}$, the saturated moisture content $\theta_{\text{s}}$, the residual moisture content $\theta_{\text{r}}$, and the curve fitting parameters ($\alpha$, $n$) constitute the hydraulic parameters used in this work.

Having obtained the elevation information and soil hydraulic parameters, these attributes are combined to form the primary data set for delineation. However, since the elevation values and soil hydraulic parameters have different units and magnitudes, the primary data-set contains features with highly variable magnitudes. This can cause some features to have a greater influence on the clustering process than others, leading to biased results. To avoid this, the data-set is normalized so as to bring all the features to the same magnitude levels. Normalization also ensures that each feature contributes equally to the clustering process. This is particularly important when using distance-based clustering methods like \textit{k}-means. The elbow method is then used to determine the optimal number of clusters or management zones. Once the optimal number of clusters is determined, clustering is performed on the normalized data using the k-means method to obtain the different management zones.

 K-means clustering is a popular unsupervised machine learning algorithm used to partition a given dataset into K clusters. It works by iteratively assigning data points to their nearest cluster center (centroid) based on a distance metric, and then updating the centroids based on the mean of the assigned data points. The algorithm is as follows:
\begin{enumerate}
	\item Choose the number of clusters, K, and randomly initialize K centroids.
	\item Assign each data point to the nearest centroid based on a distance metric (usually Euclidean distance).
	\item Update each centroid to be the mean of the data points assigned to it.
	\item Repeat Steps (b) and (c) until convergence, i.e., when the centroids no longer change significantly or a maximum number of iterations is reached.
\end{enumerate}
% Each management zone covers a section of the field with comparable hydraulic parameters and elevation.
The main steps involved in this stage of the delineation process are summarized in Figure~\ref{fig:delin_stage_2}.

\begin{figure}
    \centering
    {\includegraphics[width=0.8\textwidth]{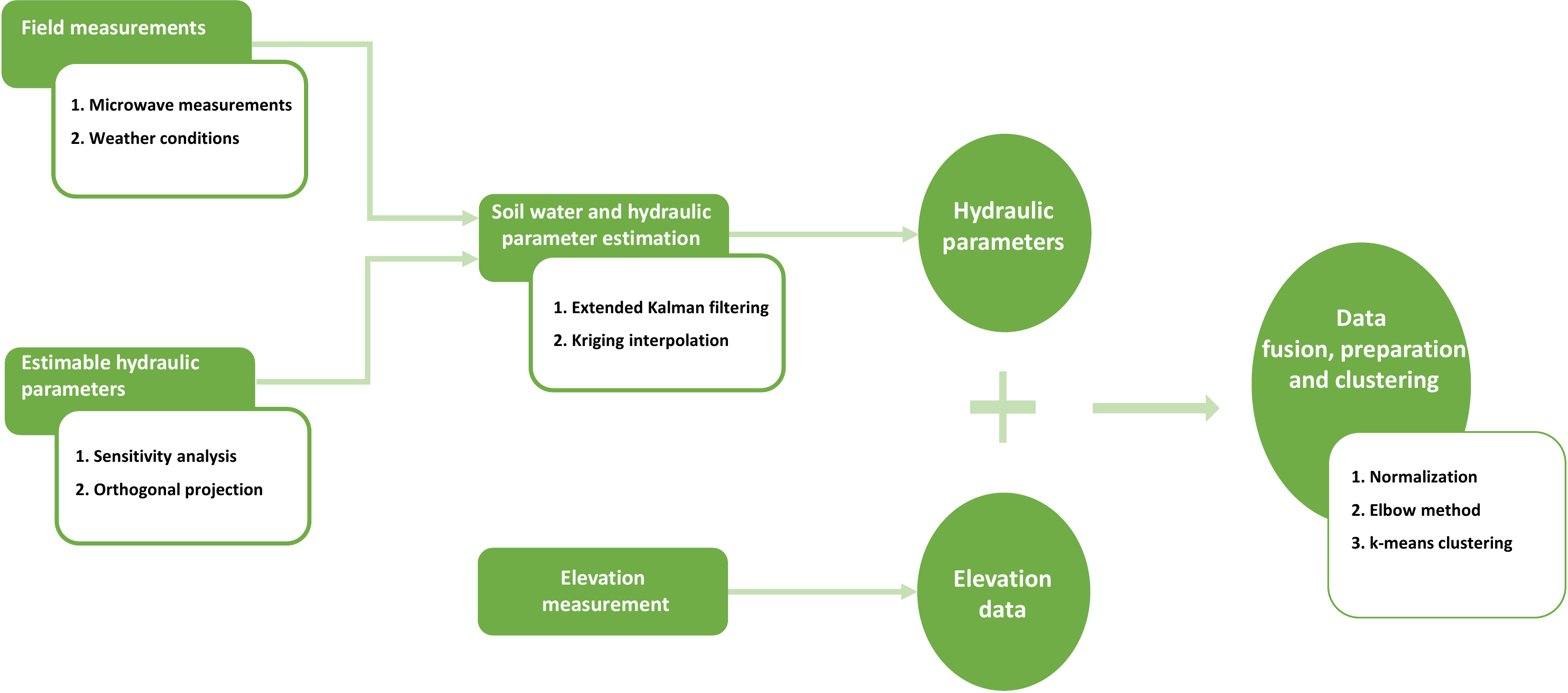}}
    \caption{An illustration of Stage 2 of the proposed delineation process.}
    \label{fig:delin_stage_2}
\end{figure}

\item \textbf{Stage 3: mapping homogeneous areas to irrigation equipment resolution.}  After clustering in the second stage, the resulting management zones are often irregularly shaped, which makes it difficult to apply water efficiently using the irrigation equipment. To overcome this challenge, the management zones are mapped to the resolution of the field's irrigation equipment by breaking them up into smaller, more uniform sub-zones. This ensures that the sub-zones align with the capabilities of the equipment. By aligning the management zones with the equipment's capabilities, the irrigation can be applied more accurately and efficiently, ensuring that water is only applied where it is needed. This mapping step is critical in realizing the full benefits of the delineation process, as it enables more precise and efficient application of irrigation water based on the specific requirements of crops in each sub-zone.
\end{enumerate}

\subsection{Supervised learning for soil moisture modelling}

\subsubsection{Soil moisture modeling in the management zones}
By using elevation data to define management zones, the soil profile within each zone is anticipated to have less geospatial variability. This allows the use of the 1D Richards equation to describe the dynamics of soil moisture in each management zone. The 1D Richards equation can be expressed in capillary pressure head form as:
\begin{equation}\label{eq:RE_1D}
  c(\psi)\frac{\partial \psi}{\partial t} = \frac{\partial}{\partial z}\bigg[K(\psi)\bigg(\frac{\partial \psi  }{\partial z}+ 1\bigg)\bigg]-\rho(\psi)\mathcal{R}\left(\text{K}_{\text{c}}, \text{ET}_0, \text{z}_{\text{r}}\right)
\end{equation}
In Equation~\ref{eq:RE_1D}, $\psi~(m)$ is the capillary pressure head, which describes the status of water in soil, $t~(s)$ represents time, $z(m)$ is the spatial coordinate, $K(\cdot)~(m \cdot s^{-1})$ is the unsaturated hydraulic water conductivity, $ c(\cdot)~(m^{-1})$ is the capillary capacity. $K(\cdot)$ and $ c(\cdot)$ are parameterized by models of Maulem and van Genuchten.
$\rho(\psi)~(-)$ is a dimensionless stress water factor, $\mathcal{R}(\cdot)$ is the root water uptake model which is a function of the crop coefficient $\text{K}_{\text{c}}~(-)$, the reference evapotranspiration $\text{ET}_0 ~(m \cdot s^{-1})$, and the rooting depth $\text{z}_{\text{r}}~(m)$. Note that the hydraulic parameters obtained in the center of each management zone are used to parameterize the Richards equation.
The 1D Richards equation is solved numerically for the following boundary conditions:
\begin{align}
    &\frac{\partial (\psi+z)}{\partial z}\bigg|_{z=H_z}=1\label{eq:botBC} \\
    &\frac{\partial \psi}{\partial z}\bigg|_{z=0}=-1-\frac{u^{\text{irrig } } - \text{EV}}{K(\psi)}\label{eq:topBC}
\end{align}
where $H_z~(m)$, $u^{\text{irrig}}~(m \cdot s^{-1})$, and $\text{EV}~(m \cdot s^{-1})$ in Eqs. (\ref{eq:botBC}) and (\ref{eq:topBC})  represent the depth of the soil column, the irrigation rate and the evaporation rate, respectively. The relationship between the pressure head $\psi$ and the volumetric moisture content $\theta_v$ is given as:
\begin{eqnarray}\label{eq:thetareln}
\theta_{\text{v}} (\psi)=\theta_r +(\theta_{\text{s}}-\theta_{\text{r}})\bigg[\frac{1}{1+(-\alpha \psi)^n}\bigg]^{1-\frac{1}{n}}
\end{eqnarray}
where the parameters in Equation~\ref{eq:thetareln} are as defined in Section~\ref{delination}. It is important to add that the numerical approach proposed in~\cite{agyeman2023lstm} is employed to solve Equation~\ref{eq:RE_1D}. Interested readers may refer to the aforementioned reference for a detailed explanation of the numerical approach. After applying the numerical method, the 1D Richards equation can be written in state-space form as:
 \begin{align}
x_{k+1}&=\mathcal{F}(x_k,u_k, \bm{\theta}) + \omega_k \label{eq:state_equation}\\
y_{k}&=\mathcal{H}(x_k,\bm{\theta}) \label{eq:output_equation}
\end{align}
where $x_k\in \mathbb{R}^{N_x}$ represents the state vector containing $N_x$ capillary pressure head values for the corresponding spatial nodes. $u_k$ represents the input vector containing the irrigation amount, precipitation, daily reference evapotranspiration, the crop coefficient, and the rooting depth. $\omega_k$ is the model disturbance, and $\bm{\theta}$ represents the 5 hydraulic parameters. The volumetric water content $\theta_{\text{v}}$ is chosen as the output $y_k$. Consequently, $y_k\in \mathbb{R}^{N_y}$ represents the state vector containing $N_p = N_x$ volumetric water content values for the corresponding spatial nodes.

\subsubsection{LSTM network development} \label{lstm_mod_dev}
Mechanistic agro-hydrological models like the Richards equation present practical challenges when used in the design of model-based irrigation schedulers. One of the significant issues is their numerical complexity, which  renders the resulting scheduler computationally inefficient. To overcome this drawback, an alternative solution is proposed, which involves using an LSTM network to model the dynamics of volumetric soil moisture in each management zone.

Long Short-Term Memory (LSTM) is a specialized type of Recurrent Neural Network (RNN) that can effectively capture long-term dependencies. Unlike conventional RNNs, which only have a hidden state ($h$), an LSTM memory block comprises four elements: a cell state ($C$), an input gate ($i$), an output gate ($o$), and a forget gate ($f$). The cell state is responsible for storing information over extended periods of time, while the gates, which are made up of a sigmoid activation function ($\sigma$) and a pointwise multiplication operation ($\otimes$), regulate the information flow into and out of the cell state. The three gates each act as a controller of the information within the cell state.
The LSTM evaluates the mapping of the input sequence $m_k,~k=1,...,T$, where $T$ is the length of the input sequence, to the predicted output sequence $\hat{y}_k$ by iterating through the following equations:
\begin{gather}
	i_k=\sigma(w_im_k + U_ih_{k-1} + b_i)\\
	f_k=\sigma(w_fm_k + U_fh_{k-1} + b_f)\\
	o_k=\sigma(w_om_k + U_oh_{k-1} + b_o)\\
	\tilde{C}_k = \text{tanh}(w_cm_k + U_ch_{k-1} + b_c)\\ 
	C_k = f_k\otimes C_{k-1} + i_k\otimes \tilde{C}_k\\
	h_k = o_k\otimes \text{tanh}(C_k)\\
	\hat{y}_k = w_yh_k+ b_y\label{eq:final}
	%\hat{x}_k = w_yh_k+ b_y\label{eq:final}
\end{gather}
where $w_i~, w_f, w_o$ are the weights for the input, forget, and output gates to the input, respectively. $U_i,~U_f,~U_o$ are the matrices of the weights from the input, forget, and output gates to the hidden states, respectively. $b_i,~b_f,~b_o$ are the bias vectors associated with the input, forget, and output gates. The predicted output from the LSTM is calculated using Equation (\ref{eq:final}) where $w_y$ and $b_y$ denote the weight matrix and bias vector for the output, respectively.

The LSTM network is developed using a dataset obtained by performing open-loop simulations of the calibrated 1D Richards equation for each management zone. This dataset is generated by solving Equations~(\ref{eq:state_equation}) and (\ref{eq:output_equation}) for randomly generated inputs and initial states $x_0$, with some level of noise included in the simulations to account for model uncertainty and to enhance the generalization ability and robustness of the LSTM.

For irrigation scheduling purposes, it is sufficient to focus on the soil moisture dynamics in the root zone of the management zone. In this study, a weighted sum of the $N_y$ volumetric moisture content values of  Equation~\ref{eq:output_equation} is used to characterize the root zone volumetric moisture content. Specifically, a weight of 40\% is assigned to the average volumetric moisture content value in the upper quarter of the rooting depth $\text{z}_{\text{r}}$, a weight of 30\% is assigned to the average volumetric moisture content value in the second quarter, a weight of 20\% is assigned to the average volumetric moisture content value in the third quarter, and a weight of 10\% is assigned to the average volumetric moisture content value in the last quarter. This arrangement is consistent with the relative amount of moisture extracted from different depths of the crop rooting depth. The proposed approach for predicting the root zone soil moisture leverages the universal approximation ability of neural networks and directly predicts the weighted average of the soil moisture at various depths of the root zone using a single LSTM model.

\begin{figure}
    \centering
    {\includegraphics[width=0.8\textwidth]{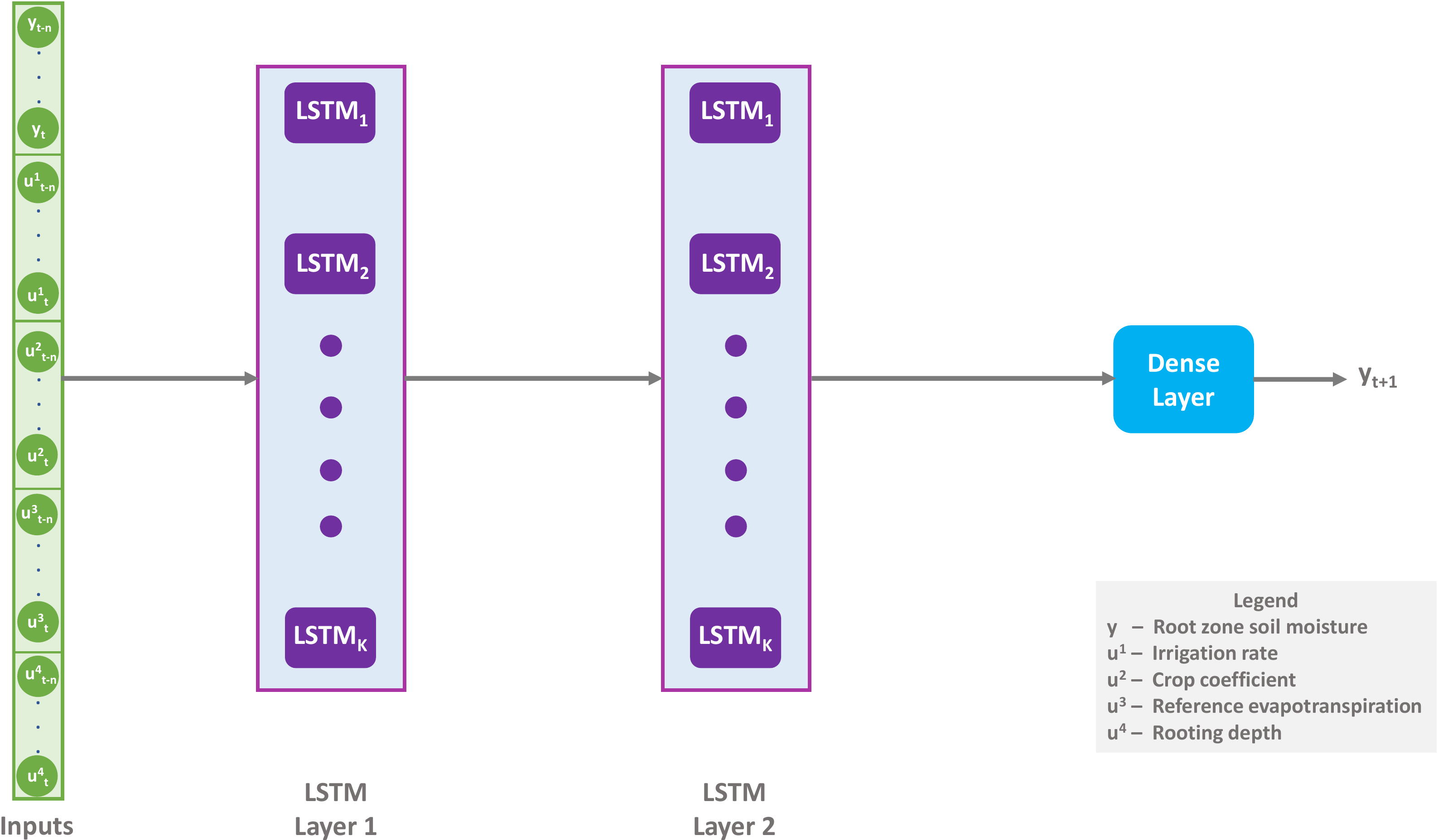}}
    \caption{A schematic diagram of the proposed LSTM model for each management zone.}
    \label{fig:lstm_diag}
\end{figure}

The LSTM model is designed to predict the one-day-ahead root zone volumetric water content $\hat{y}_{t+1}$ using current and past inputs, including current and past root zone volumetric water content $\hat{y}(t=0,...,l)$, precipitation (i.e. rain and irrigation rate) $u^{\text{irrig}} (t=0,...,l)$, crop coefficient values $\text{K}_{\text{c}} (t=0,...,l)$, reference evapotranspiration values $\text{ET}_0 (t=0,...,l)$, and the current and  past rooting depth values $\text{z}_{\text{r}} (t=0,...,l)$. The time lag or sequence length used for model development is denoted as $l$. The proposed approach incorporates the rooting depth as one of the inputs to the LSTM model, which allows for the time-varying nature of the rooting depth phenomenon to be considered. This integration has the potential to improve irrigation and water use efficiency. A schematic representation of the proposed LSTM network architecture is shown in Figure~\ref{fig:lstm_diag}.

 The root mean squared error (RMSE) and the coefficient of determination (R$^2$) are used to evaluate the predictive capability of the identified LSTM model for each management zone. The RMSE is used to quantify the prediction error in the units of the volumetric moisture content and it is mathematically defined as:
\begin{equation}\label{rmse}
	\text{RMSE} = \left[\frac{\sum_{k=1}^n(y_k - \hat{y}_k)^2}{n}\right]^{\frac{1}{2}}
\end{equation}
where $y_k$ and $\hat{y}_k$ represent the true and predicted values of the root zone soil moisture content for day $k$. The R$^2$ is used to quantify the ability of the identified LSTM model to explain the variance in the observed data and it is expressed as follows:
\begin{equation}
	\text{R}^2 =  1 - \frac{\sum_{k=1}^n(y_k - \hat{y}_k)^2}{\sum_{k=1}^n(y_k - \bar{y}_k)^2}  
\end{equation}
where $\bar{y}_k$ represents the mean of the true root zone soil moisture content values.

\subsection{Mixed-integer MPC for irrigation scheduling}
The irrigation scheduling problem is modeled as a variant of the warehouse location problem, where the irrigation implementing equipment serves as the warehouse, and the decision to operate the equipment is modeled as the decision to operate the warehouse. The fixed cost of operating the equipment is modeled as the fixed cost of opening a warehouse, and the variable cost of supplying water to the fields is modeled as the variable cost of supplying the warehouse. The objective function is to minimize the total cost of operating the irrigation equipment over the planning horizon, subject to constraints that ensure that the demand for water in each management zone is met, and that the total supply of water does not exceed the capacity of the irrigation equipment. To modify the formulation to maintain the root zone soil moisture within a target zone, a penalty cost for violating the target zone is added to the objective function, and constraints are updated to ensure that the root zone soil moisture is maintained within the target zone. Furthermore, the dynamics of root zone soil moisture must be considered during the planning horizon, as the root zone soil moisture level changes over time. This modified formulation results in a mixed-integer MPC problem with zone objectives (zone control), where the identified LSTM model for each management zone is employed as the dynamic model in the formulation. Additionally, a soft constraint approach is employed to realize the objective of maintaining the root zone soil moisture in the target zone where slack variables are introduced in the formulation to relax the bounds of the target zone. By solving the mixed-integer MPC with zone objectives problem, the optimal schedule for operating the irrigation equipment can be determined that minimizes the total cost of operation while maintaining the root zone soil moisture within the target zone throughout the planning horizon.
For day $d$ and a planning horizon of $N$ days, the scheduler $\mathbb{P}(y)$ is formulated for a single management zone as follows:
\begin{subequations}
	\begin{alignat}{2}
		\min_{\bm{y},~\bm{\bar{\epsilon}},~\bm{\underline{\epsilon}},~\bm{u}^{\text{irrig}},~\bm{c}} ~ &\sum_{k=d+1}^{d+N}\left[\bar{Q}\bar{\epsilon}^2_k + \underline{Q}\underline{\epsilon}^2_k \right] + \sum_{k=d}^{d+N-1}R_cc_k &&+ \sum_{k=d}^{d+N-1}R_uu_k^{\text{irrig}} \label{eq:obj} \\
		\notag
		&\textrm{s.t. }&&\\ 
		&y_{k+1} = \mathcal{F}(\{{\gamma}\}_{k-l}^{k},\eta) &&k\in [d,d+N-1] \label{eq:cons1}\\
		&y_d = y(d) \label{eq:cons2} \\
		&\underline{\nu} - \underline{\epsilon}_k\leq y_k \leq \bar{\nu} + \bar{\epsilon}_k, && k\in [d+1,d+N] \label{eq:cons3} \\
		&c_k\underline{u}^{\text{irrig}} \leq u_k^{\text{irrig}} \leq c_k\bar{u}^{\text{irrig}}, && k\in [d,d+N-1] \label{eq:cons4} \\
		&c_k= \{0, 1\}, && k\in [d,d+N-1] \label{eq:cons5}\\ 
		&\underline{\epsilon}_k \geq 0, \quad \bar{\epsilon}_k\geq 0, && k\in [d+1,d+N] \label{eq:cons6} 
	\end{alignat}
\end{subequations}
where $k \in \mathbb{Z}^+$, $\bm{y}\coloneqq [ y_d, y_{d+1},...,y_{d+N}]$, $\bm{\bar{\epsilon}}\coloneqq [ \bar{\epsilon}_{d+1}, \bar{\epsilon}_{d+2},...,\bar{\epsilon}_{d+N}]$, $\bm{\underline{\epsilon}}\coloneqq [ \underline{\epsilon}_{d+1}, \underline{\epsilon}_{d+2},...,\underline{\epsilon}_{d+N}]$, $\bm{c}\coloneqq [ c_{d}, c_{d+1},...,c_{d+N-1}]$, $\bm{u}^{\text{irrig}}\coloneqq [ u_{d}^{\text{irrig}}, u_{d+1}^{\text{irrig}},...,u_{d+N-1}^{\text{irrig}}]$, and $\{\gamma\}_{k-l}^{k} \coloneqq [ \gamma_{k-l}, \gamma_{k-l-1},..,\gamma_{k}]$ where $\gamma\in[y,\text{K}_{\text{c}}, \text{ET}_0, u^{\text{irrig}}, \text{z}_{\text{r}}]$. 
$\underline{\epsilon}_k$ and $\bar{\epsilon}_k$~(\ref{eq:obj}, \ref{eq:cons6}) are nonnegative slack variables that are introduced to relax the bounds of the target zone $(\underline{\nu}_k,\bar{\nu}_k)$ in~(\ref{eq:cons3}). The per-unit costs associated with the violation of the lower and upper bounds of the target zone are represented with  $\underline{Q}$ and $\bar{Q}$, respectively. $R_u$ and $R_c$ represent the per-unit cost of the applied irrigation amount and the fixed cost associated with operating the irrigation implementing equipment. The binary decision variable $c$ encodes the daily irrigation decision.
% $\underline{Q}$ and $\bar{Q}$ are the per-unit costs associated with the violation of the lower and upper bounds of the target zone, respectively. $R_c$ is the fixed cost associated with the operation of the irrigation implementing system, and $R_u$ is the per-unit cost of the irrigation rate $u^{\text{irrig}}$. The binary variable ($c$) encodes the daily discrete irrigation decision. 
The cost function~(\ref{eq:obj}) incorporates the objectives of maintaining the root zone volumetric moisture content in a target zone in order to ensure optimal water uptake in crops by minimizing the violation of the target zone $\sum_{k=d+1}^{d+N}\left[\bar{Q}\bar{\epsilon}^2_k + \underline{Q}\underline{\epsilon}^2_k \right]$, minimizing the irrigation cost $\sum_{k=d}^{d+N-1}R_cc_k$, and minimizing the irrigation amount $\sum_{k=d}^{d+N-1}R_uu_k^{\text{irrig}}$. Constraint~(\ref{eq:cons1}) corresponds to the LSTM model of the root zone volumetric moisture content, and $\eta$ in this constraint represents the weights and bias terms associated with the identified LSTM model. In this work, the initial state is represented with Constraint~(\ref{eq:cons2}). Its value is assumed to be known and during the evaluation of the scheduler, it is chosen to be the estimated state obtained after the state and parameter estimation step outlined in Section \ref{delination} has converged.  Constraint~(\ref{eq:cons4}) refers to the maximum amount of water that can be supplied during an irrigation event on day $k$. When the irrigation decision on day $k$ is ``a no decision" ($c_k=0$), Constraint~(\ref{eq:cons4}) requires that the irrigation rate must be set to 0. Conversely, when the irrigation decision on a particular day is ``a yes decision" ($c_k=1$), Constraint~(\ref{eq:cons4}) specifies that the prescribed irrigation rate must not exceed its upper bound ($\bar{u}^{\text{irrig}}$) and must be at least as large as its lower bound ($\underline{u}^{\text{irrig}}$).

After providing a detailed description of the scheduler design for a single management zone, the daily scheduler $\mathbb{P}_M(y)$ for a field with $M$ management zones is formulated as follows:
\begin{subequations}
	\begin{alignat}{3}\label{eq:obj_sv}
		\min_{\bm{Y, ~\bar{E}, ~\underline{E},~ U^{\text{irrig}},~ c}} ~ &\sum_{j=1}^{M} \sum_{k=d+1}^{d+N}\left[\bar{Q}(j)\bar{\epsilon}^2_{jk} +\underline{Q}(j)\underline{\epsilon}^2_{jk} \right] &&+R_c\sum_{k=d}^{d+N-1}c_{k} &&+ R_u\sum_{j=1}^{M}\sum_{k=d}^{d+N-1}u_{jk}^{\text{irrig}}\\
		\notag
		&\textrm{s.t. }&& ~ &&\\
		&y_{j(k+1)} =\mathcal{F}^j(\{\gamma_j\}_{k-l}^{k},\eta_j) &&j\in \tilde{M}, &&k\in [d,d+N-1]\\
		&y_{j0} = y_j(0), && j\in \tilde{M} && \\
		&\underline{\nu}_j - \underline{\epsilon}_{jk} \leq y_{jk}\leq \bar{\nu}_j + \bar{\epsilon}_{jk}, &&j\in \tilde{M} ,&&k \in [d+1,d+N]\\
		&c_{k} \underline{u}_{j}^{\text{irrig}}\leq u_{jk}\leq c_{k} \bar{u}_{j}^{\text{irrig}}, && j\in \tilde{M} , &&k \in [d,d+N-1]\\
		&c_{k}\in \{0, 1\}, && ~&& k \in [d,d+N-1]\\
		&\underline{\epsilon}_{jk} \geq 0, \quad \bar{\epsilon}_{jk}\geq 0, &&j\in \tilde{M} ,&&k \in [d+1,d+N] \label{eq:lastconst_sv}
	\end{alignat}
\end{subequations}
where $j \in \mathbb{Z}^+$, $k \in \mathbb{Z}^+$, $\bm{Y} \coloneqq [\bm{y}_1,\bm{y}_2,....,\bm{y}_M ]$, $\bm{\bar{E}} \coloneqq [\bm{\bar{\epsilon}}_1,\bm{\bar{\epsilon}}_2,....,\bm{\bar{\epsilon}}_M ]$, $\bm{\underline{E}} \coloneqq [\bm{\underline{\epsilon}}_1,\bm{\underline{\epsilon}}_2,....,\bm{\underline{\epsilon}}_M ]$, and $\bm{U}^{\text{irrig}} \coloneqq [\bm{u}_1^{\text{irrig}},\bm{u}_2^{\text{irrig}},....,\bm{u}_M^{\text{irrig}} ]$. $\tilde{M}$ is a closed set that contains the $[j_1, j_2, ..., j_M]$ positional indices of the $M$ management zones.
Essentially, the formulation $\mathbb{P}_{\text{M}}(y)$ aims to duplicate the formulation $\mathbb{P}(y)$ for each of the $M$ management zones present in the field. Furthermore, it is apparent that in all $M$ management zones, the binary values specified for the planning horizon remain the same in $\mathbb{P}_{\text{M}}(y)$.

In general, solving $\mathbb{P}_{\text{M}}(y)$ provides the irrigation decision and rates for all $M$ management zones in the field. However, solving $\mathbb{P}_{\text{M}}(y)$ is expected to be more challenging than $\mathbb{P}(y)$, and determining its tuning parameters is also expected to be difficult. If there is a way to determine the irrigation decision sequence without explicitly solving $\mathbb{P}_{\text{M}}(y)$, $M$ copies of $\mathbb{P}(y)$ can be solved in parallel instead of $\mathbb{P}_{\text{M}}(y)$. In the following sections, we explain how the concept of a limiting management zone and reinforcement learning technique can be used to identify a binding irrigation sequence for the $M$ management zones, which allows for solving $M$ copies of $\mathbb{P}(y)$ instead of $\mathbb{P}_{\text{M}}(y)$.

\subsection{Reinforcement Learning for Irrigation Decision Determination}\label{agent_development}
Reinforcement learning (RL) is a feedback-driven machine learning approach that focuses on an agent's ability to learn optimal decision-making strategies through trial and error within an environment. In RL, an agent interacts with the environment, which can be physical or virtual, by taking actions based on the current state or observation and receiving feedback in the form of rewards or penalties. The RL framework is typically formulated as a Markov decision process (MDP), consisting of several key elements. First, there is a set of states or observations (denoted as $x^{rl}$), representing the different configurations of the environment. The agent perceives these states to make informed decisions. Additionally, there exists a set of actions ($u^{rl}$) that the agent can take in response to the observed states.  The environment's dynamics are captured by a transition function or dynamics function ($f^{rl}$), which describes how the states transition from one to another based on the chosen action. Crucially, RL introduces the concept of rewards ($r$), which represent the positive or negative feedback received by the agent for each state-action pair. Rewards serve as a guiding signal for the agent to evaluate the desirability of its actions. The agent aims to maximize the cumulative reward it receives over multiple interactions with the environment. To achieve this, the agent learns a policy ($\pi(u|x)$), which is a mapping from states to actions.

To determine the optimal irrigation decision sequence for the proposed scheduler, we employ the reinforcement learning (RL) paradigm. The objective is to train an agent capable of formulating $\mathbb{P}(y)$ for each of the $M$ management zones within $\mathbb{P}_{\text{M}}(y)$. The goal is to learn a policy that can accurately calculate the daily irrigation decision and rate, enabling the scheduler to achieve its objectives for each management zone. During this process, the RL agent interacts with a calibrated 1D Richards equation, which serves as a model for capturing the dynamics of soil moisture within each management zone. It is important to note that the transition dynamics used for training the RL agent are determined by Equations (\ref{eq:state_equation}) and (\ref{eq:output_equation}). For each management zone, the policy takes input parameters such as the $N_y$ volumetric moisture content values derived from Equation \ref{eq:output_equation}, reference evapotranspiration, crop coefficient, and rooting depth at the current time instant $k$. These parameters can be summarized as $x^{rl}_k = [y_k,~\text{ET}_k,~K_{c_k},~z_{r_k}]$. Using this information, the policy then determines actions $u_k^{rl} = [c_k~u_k^{\text{irrig}}]$, that is the daily irrigation decision $c_k$ and the daily rate $u_k^{\text{irrig}}$, that maximize a specified reward function. The employed reward function $r_k$ is defined as the negative of the cost function in $\mathbb{P}(y)$. Its purpose is to guide the RL agent towards decisions that minimize costs and achieve desired outcomes, and it is defined as:

\begin{align}
 r_k &= -R^{\text{Z}}_k -R_cc_k -  R_uu_k^{\text{irrig}} \label{eq:reward_a}\\ 
 R^{\text{Z}}_k &= \begin{cases}
\underline{Q}\times|\theta^{\text{RZ}}_k - \underline{\nu}| & \text{if } \theta^{\text{RZ}}_k < \underline{\nu}\\
\overline{Q}\times|\theta^{\text{RZ}}_k - \overline{\nu}| & \text{if } \theta^{\text{RZ}}_k > \overline{\nu}\\ 
0 & \text{if } \underline{\nu} \leq \theta^{\text{RZ}}_k \leq \overline{\nu}\\
\end{cases} \label{eq:reward_b}
\end{align}
In Equations (\ref{eq:reward_a}) and (\ref{eq:reward_b}), $R_c$, $R_u$, $\underline{Q}$, $\overline{Q}$, $\underline{\nu}$, and $\overline{\nu}$ have the same definitions as in the scheduler formulation of $\mathbb{P}(y)$. $\theta_k^{\text{RZ}}$ represents the root zone soil moisture at time step $k$, which is the weighted average of the $N_y$ volumetric moisture content values of Equation (\ref{eq:output_equation}). The calculation of $\theta_k^{RZ}$ is outlined in Section \ref{lstm_mod_dev}. The primary objective of the RL agent is to maintain the root zone soil moisture within the specified target range while minimizing both total water consumption and irrigation cost. As a consequence, the agent incurs a penalty of $\underline{Q}\times|\theta^{\text{RZ}}_k - \underline{\nu}|$ when its actions result in the root zone soil moisture content falling below the lower threshold $\underline{\nu}$. Similarly, the agent is penalized with $\overline{Q}\times|\theta^{\text{RZ}}_k - \overline{\nu}|$ if its actions cause the root zone soil moisture content to exceed the upper threshold $\overline{\nu}$. When the agent successfully maintains the root zone soil moisture within the desired range, it receives a reward of 0. Additionally, there are penalties imposed for specific actions taken by the agent. A penalty of $R_cc_k$ is applied when the agent decides to perform irrigation (indicated by a ``yes" decision) and a penalty of $R_uu_k^{\text{irrig}}$ is incurred for a non-zero irrigation rate. Overall, the reward function is designed to guide the agent towards making irrigation decisions that balance the objectives of maintaining optimal soil moisture levels, minimizing water usage, and reducing irrigation costs.

\begin{figure}
    \centering
    {\includegraphics[width=0.8\textwidth]{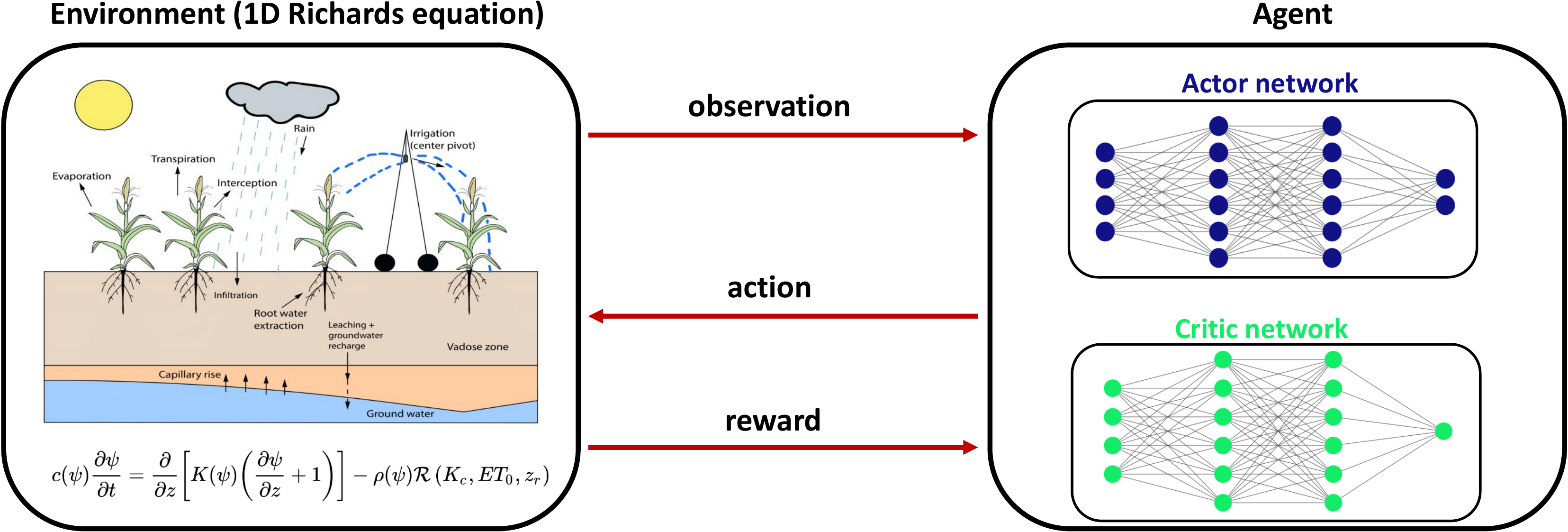}}
    \caption{Interaction between environment and agent in reinforcement learning.}
    \label{fig:lstm_diag_1}
\end{figure}

The Proximal Policy Optimization (PPO) algorithm, which is utilized in this study, is an actor-critic method designed to optimize policies for all reinforcement learning (RL) agents. PPO aims to find the optimal policy by simultaneously estimating the value function (critic) and the policy (actor) for the agent. By employing an actor-critic architecture, PPO leverages the value function to estimate the expected return and guides the policy updates based on the estimated advantages. This combination allows PPO to learn an effective policy that maximizes the cumulative reward signal over time. During the training process, PPO collects experience data by having the agent interact with the environment and uses this data to update both the policy and value function. The policy update is performed by minimizing a surrogate loss function, which approximates the policy's performance in terms of the expected reward. This update process is guided by the estimated advantages provided by the value function. PPO further enhances stability in the learning process by using a clipped objective function, which restricts policy updates to prevent large deviations from the current policy. This clipping mechanism ensures a more controlled policy update while maximizing the agent's learning progress. PPO's actor-critic approach and its ability to balance exploration and exploitation make it a well-established and effective algorithm for various RL tasks, making it a suitable choice for this study. For a more detailed understanding of PPO, interested readers may refer to \cite{schulman2017proximal}. Figure~\ref{fig:lstm_diag_1} illustrates a diagrammatic representation of the agent-environment interaction during training.

\subsection{Binding irrigation decision determination and synchronized irrigation management algorithm}
In Section~\ref{agent_development}, we discussed the decentralized training approach utilized for the RL agents in the $M$ management zones. As a result, when evaluating the policies of these trained agents for a planning horizon of $N$ on a specific day $d$, it is expected that the irrigation decision sequences obtained for each management zone would differ. However, adhering to formulation $\mathbb{P}_{\text{M}}(y)$ necessitates utilizing the same irrigation decision sequence for all $M$ management zones during evaluation. To address this requirement, we introduce the concepts of the limiting irrigation management zone and the binding irrigation decision sequence. The limiting irrigation management zone is determined as the $j^{\text{th}}$ management zone in $M$ that exhibits the earliest non-zero irrigation decision within the planning horizon. Its irrigation decision sequence is then chosen as the binding irrigation sequence. By selecting the irrigation decision sequence of the limiting management zone as the binding irrigation decision, the scheduler can apply irrigation in a timely manner to all management zones, effectively avoiding crop stress and maintaining the integrity of the lower bound of the target zone—a factor of utmost importance. The binding irrigation sequence replaces the irrigation sequence in each management zone, ensuring that the same irrigation sequence is enforced across all management zones, in accordance with formulation $\mathbb{P}_{\text{M}}(y)$.

An illustrative example can help clarify the concepts of the limiting irrigation management zone and binding irrigation sequence. Let us consider a field divided into three management zones:  $\text{MZ}_1$, $\text{MZ}_2$, and $\text{MZ}_3$. Each zone has its own RL agent with a designated policy: $\pi_{\text{MZ}_1}(\cdot)$, $\pi_{\text{MZ}_2}(\cdot)$, and $\pi_{\text{MZ}_3}(\cdot)$. On a given day, denoted as $d$, the observations within each management zone are represented as $x^{rl}_{\text{MZ}_1}$, $x^{rl}_{\text{MZ}_2}$, and $x^{rl}_{\text{MZ}_3}$. It is assumed that the agents are evaluated for a planning horizon of seven days. After evaluating the agents, we obtain the following irrigation sequences ($\bm{c}$) for each management zone:  $\bm{c}_{\text{MZ}_1} =\pi_{\text{MZ}_1}(\bm{c}| x^{rl}_{\text{MZ}_1}) = [0,1,0,1,0,0,0]$, $\bm{c}_{\text{MZ}_2} =\pi_{\text{MZ}_2}(\bm{c}| x^{rl}_{\text{MZ}_2}) = [0,0,0,1,0,0,1]$, and $\bm{c}_{\text{MZ}_3} =\pi_{\text{MZ}_3}(\bm{c}| x^{rl}_{\text{MZ}_3}) = [0,0,0,0,1,0,1]$. Based on the irrigation sequences obtained, it is evident that $\text{MZ}_1$ exhibits the earliest non-zero irrigation decision among all the management zones. Consequently, $\text{MZ}_1$ is identified as the limiting irrigation management zone. In the provided example, the irrigation sequence for $\text{MZ}_1$ signifies that irrigation is initiated on the second day and repeated on the fourth day. In contrast, $\text{MZ}_2$ and $\text{MZ}_3$ have their first non-zero irrigation decisions occurring on subsequent days, with $\text{MZ}_2$ commencing irrigation on the fourth day and $\text{MZ}_3$ on the fifth day. By designating $\text{MZ}_1$ as the limiting management zone, it is established that the irrigation decisions made within $\text{MZ}_1$ serve as the reference for irrigation scheduling in the other zones. The binding irrigation sequence of $\text{MZ}_1$, $[0, 1, 0, 1, 0, 0, 0]$, is enforced in $\text{MZ}_2$ and $\text{MZ}_3$, ensuring synchronization and timely irrigation across all management zones. This coordinated approach guarantees that each zone receives irrigation in a prompt manner, mitigating the risk of crop stress and preventing violations of the lower bound of the target zone. 
% By adhering to the irrigation schedule dictated by $\text{MZ}_1$, the entire field's irrigation management is optimized, promoting crop health and productivity.

It is noteworthy that while the determination of the irrigation decision sequence in each management zone can be regarded as optimal, the selection of the binding irrigation decision sequence is deemed suboptimal. This is due to the fact that the binding sequence is chosen based on the earliest non-zero irrigation decision observed in the limiting irrigation management zone, rather than being derived through a comprehensive optimization process considering the entire field. Nevertheless, this suboptimal approach is essential to prevent crop stress and violations of the lower bound of the target zone. Additionally, by employing the suboptimal approach to determine the binding irrigation decision sequence, a uniform sequence can be established and implemented across all management zones of the field. This simplifies the overall optimization problem, as it allows for solving multiple copies (M) of the problem $\mathbb{P}(y)$ in parallel, instead of solving the problem $\mathbb{P}_{\text{M}}(y)$ that encompasses the entire field. Parallel execution of these copies capitalizes on the multicore architecture of modern CPUs, distributing each copy to a separate processor core, resulting in substantial reductions in computation time. Parallel computing is a widely adopted strategy for tackling computationally intensive optimization problems, offering notable advantages in terms of efficiency and scalability of the optimization algorithm. By leveraging parallel computing, the irrigation management system can achieve significant improvements in computational speed, facilitating the timely generation of irrigation decision sequences while maintaining system performance within desired bounds.

The utilization of the pre-determined binding irrigation decision sequence in solving $M$ copies of $\mathbb{P}(y)$ transforms the mixed-integer problem into a nonlinear program, simplifying the computational aspect of the optimization problem. This transformation reduces the complexity of the problem and facilitates its efficient solution. Additionally, the outcomes obtained from evaluating the RL agents for each management zone can be leveraged to expedite the solution process of the M copies of $\mathbb{P}(y)$. Specifically, the computed irrigation rates for each zone, derived from the RL agents' results, can be employed as an initial guess during the solution of the M copies of $\mathbb{P}(y)$. Incorporating this high-quality initial guess significantly accelerates the convergence towards a solution, thereby reducing the overall computation time.

While the irrigation rates determined by the RL agent in the limiting management zone can be directly implemented as they align with the binding irrigation decision sequence across the planning horizon, solving $\mathbb{P}(y)$ for the limiting zone offers the advantage of further optimizing these irrigation rates. On the other hand, for the remaining management zones, the irrigation rates obtained from the RL agents only serve as a reliable initial guess, as they do not precisely match the binding irrigation sequence. Hence, it is necessary to solve $\mathbb{P}(y)$ for these zones to ensure that the irrigation rates align precisely with the binding irrigation sequence. For irrigation rates to be considered in line with the binding irrigation sequence, a non-zero irrigation rate should correspond to a binary value of 1, while a zero irrigation rate should align with a binary value of 0. This alignment must hold true for all the days encompassing the planning horizon. 
% By ensuring this coherence between the irrigation rates and the binding irrigation sequence, the irrigation management system can achieve a consistent and synchronized irrigation schedule across all management zones, optimizing the allocation of water resources and promoting effective crop growth.

In summary, the concepts of the limiting irrigation management zone and the binding irrigation decision sequence enable synchronized irrigation scheduling across all management zones. This coordinated approach optimizes the irrigation management system, prevents crop stress, and maintains the integrity of the lower bound of the target zone. The suboptimal selection of the binding irrigation sequence is necessary for practical implementation and allows for parallel computation of the optimization problem. Moreover, utilizing the RL agents' results as an initial guess enhances the efficiency and effectiveness of the solution process for multiple copies of the optimization problem. This synchronized irrigation management approach is summarized in Algorithm~\ref{alg:alg_1}.
\begin{algorithm}
	
	\SetKwInput{Input}{Input}
	\SetKwInput{Output}{Output}
	\SetAlgoLined
	\caption{Synchronized irrigation management algorithm}
	\label{alg:alg_1}
	\BlankLine
	\Input{Observation data $x^{rl}_{\text{MZ}_1}, x^{rl}_{\text{MZ}_2}, \ldots, x^{rl}_{\text{MZ}_\text{M}}$ for $M$ management zones}
	\BlankLine
	\Input{Policies of the agents $\pi_{\text{MZ}_1}(\cdot), \pi_{\text{MZ}_2}(\cdot), \ldots, \pi_{\text{MZ}_{\text{M}}}(\cdot)$ for $M$ management zones}
	\BlankLine
	\textbf{Evaluate agents to obtain irrigation sequences}: $\bm{c}_{\text{MZ}_1}, \bm{c}_{\text{MZ}_2}, \ldots, \bm{c}_{\text{MZ}_{\text{M}}}$\
	\BlankLine
	\textbf{Determine the limiting irrigation management zone}: $\text{MZ}_\text{limiting} \leftarrow$ Management zone with the earliest non-zero irrigation decision\
	\BlankLine
	\textbf{Set the binding irrigation sequence}: $\bm{c}_{\text{binding}} \leftarrow \bm{c}_{\text{MZ}_\text{limiting}}$\
	\BlankLine
	\For{$\text{MZ} \in \{\text{MZ}_1, \text{MZ}_2, \ldots, \text{MZ}_{\text{M}}\}$}{
		\BlankLine
		Set the irrigation sequence of $\text{MZ}$ as $\bm{c}_{\text{binding}}$\
		\BlankLine
		Solve $M$ copies of the optimization problem $\mathbb{P}(y)$ using the binding irrigation sequence $\bm{c}_{\text{binding}}$ and initial guesses derived from the RL agents' results\
	}
	\BlankLine
	\Output{Optimized irrigation rates for each management zone}
	\BlankLine
\end{algorithm}

\subsection{Triggered irrigation scheduling method}
To evaluate the benefits of the proposed irrigation scheduler, a benchmark called the triggered irrigation scheduling approach is employed. This approach, which is widely employed as one of the commonly used scheduling approaches in irrigation management, aims to maintain soil moisture within the specified bounds of a target zone. An irrigation event is triggered on a specific day, denoted as $d$, when the daily root zone moisture content, represented by $\theta^{\text{RZ}}_d$, falls below the lower bound of the target zone, denoted as $\underline{\nu}$. When such a situation occurs, the irrigation rate for that day, denoted as $u^{\text{irrig}}_{d}$, is calculated using the following formula~\cite{gu2021neural}:
\begin{equation}
	u^{\text{irrig}}_{d} = \left[(\bar{\nu} - \theta^{\text{RZ}}_d) \times \text{z}_{\text{r}}\right] - P_{d+4}
	\label{eq:irrig_rate}
\end{equation}
In Equation~\ref{eq:irrig_rate}, $\text{z}_{\text{r}}$ represents the rooting depth, while $\bar{\nu}$ corresponds to the upper bound of the target zone. $P_{d+4}$ indicates the cumulative precipitation that occurs from day $d+1$ to $d+4$. The triggered irrigation scheduling approach serves as a benchmark for evaluating the benefits and effectiveness of the proposed irrigation scheduler, leveraging its widespread application and established use in the field of irrigation management. The interested reader may refer to~\cite{gu2020irrigation} and the references therein for a thorough discussion of this scheduling approach.

\subsection{Bounds of the target zone}
So far, there has been no specific information provided regarding the lower and upper boundaries of the target zone $(\underline{\nu},~\bar{\nu})$. This section aims to address this gap by offering a description and a calculation method for determining these limits. When it comes to irrigation management, the upper limit of the target zone is commonly defined as the volumetric moisture content at field capacity ($\theta_{\text{fc}}$). Hence, the upper limit of the target zone can be expressed as follows:
\begin{equation}
	\bar{\nu} = \theta_{\text{fc}}
\end{equation}
On the other hand, the lower limit of the target zone, known as the threshold ($\theta_{\text{th}}$), is calculated using the following approach:
\begin{equation}
	\underline{\nu} =\theta_{\text{th}} =  \theta_{\text{fc}} - \left[ \text{MAD}\times(\theta_{\text{fc}} -  \theta_{\text{pwp}})\right]
	\label{eq:lower_bound}
\end{equation}
In Equation~\ref{eq:lower_bound}, $\theta_{\text{pwp}}$ represents the volumetric water content at the permanent wilting point, and MAD refers to the management allowable depletion, which indicates the fraction of the total available water that is permitted to be depleted. It is important to note that this study does not focus on precisely determining $\theta_{\text{pwp}}$ and $\theta_{\text{fc}}$. Instead, literature-reported values for these parameters, pertaining to the various soil textures found within the delineated irrigation management zones, are employed. Values of $\theta_{\text{pwp}}$ and $\theta_{\text{fc}}$ for various soil textures can be found in~\cite{huffman2012irrigation}. 
\subsection{Crop Yield}
In this work, crop yield is predicted according to the following equation~\cite{bennett2011crop}:
\begin{equation}
\label{eq:yield_eqn}
Y_a =Y_m\left[1- k_y + \left(k_y\times \frac{\text{ET}_{\text{c}}}{\text{ET}_{\text{m}}}\right)\right]
\end{equation}
where $Y_a$ is the predicted yield in (Mg ha$^{-1}$), $Y_m$ is the maximum potential yield in (Mg ha$^{-1}$),  $\text{ET}_{\text{c}}$ is seasonal crop evapotranspiration (mm), $\text{ET}_{\text{m}}$ is maximum seasonal crop evapotranspiration (mm), and $k_y$ is a crop-specific yield response factor (dimensionless). It is worth noting that this approach assumes that water is the yield-limiting factor. $\text{ET}_{\text{c}}$ is related to $\text{ET}_{\text{m}}$ as follows~\cite{feddes1982simulation}:
\begin{equation}
\text{ET}_{\text{c}} = \mathcal{K}(\theta_{\text{v}})\text{ET}_{\text{m}}
\end{equation}
where $\mathcal{K}(\cdot)$ is the water stress factor,  which is a function of $\theta_v$. $\mathcal{K}(\cdot)$ is defined as 
\begin{equation}
\mathcal{K}(\theta_{\text{v}})=\begin{cases}
0 & \text{$ \theta_{{\text{v}}_1} \leq \theta_{\text{v}}$}\\
\frac{\theta_{\text{v}} - \theta_{{\text{v}}_1}}{\theta_{{\text{v}}_2}-\theta_{{\text{v}}_1}} &\text{$ \theta_{{\text{v}}_1} \leq \theta_{\text{v}} \leq \theta_{{\text{v}}_2} $}\\
1& \text{$ \theta_{{\text{v}}_2} \leq \theta_{\text{v}}\leq\theta_{{\text{v}}_3} $}\\
\frac{\theta_{\text{v}}-\theta_{{\text{v}}_w}}{\theta_{{\text{v}}_2}-\theta_{{\text{v}}_w}}&\text{$ \theta_{{\text{v}}_w}\leq \theta_{\text{v}}\leq \theta_{{\text{v}}_2} $}
\end{cases}
\end{equation}
	where $\theta_{{\text{v}}_1}$ is the volumetric moisture at the anaerobic point, $\theta_{{\text{v}}_2}$ and $\theta_{{\text{v}}_3}$ are the volumetric moisture content values between which optimal water uptake exists, and $\theta_{{\text{v}}_w}$is  volumetric moisture content at the wilting point. In this work, optimal water uptake is considered to exist between the bounds of the target zone. Consequently, $\theta_{{\text{v}}_3} = \bar{\nu}$, $\theta_{{\text{v}}_2} = \underline{\nu}$, and $\theta_{{\text{v}}_w} =\theta_{\text{pwp}}$. $\theta_{{\text{v}}_1}$ was calculated as the volumetric moisture content corresponding to a pressure head ($\psi$) of 0.1 m~\cite{capraro2008neural}.

\subsection{Study area}
\begin{figure}[!t]
\centering
    \begin{subfigure}[b]{0.49\textwidth}
        \includegraphics[width=\textwidth]{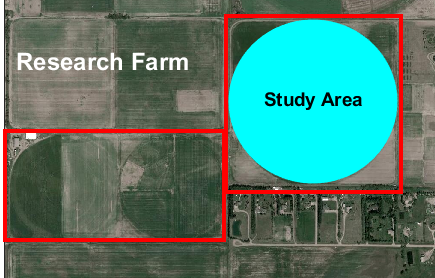}
        \caption{Layout of the study area.}
        \label{fig:layout_of_the_alberta_irrigation_center_}
    \end{subfigure}
    \hspace{6mm}
    \begin{subfigure}[b]{0.395\textwidth}
        \includegraphics[width=\textwidth]{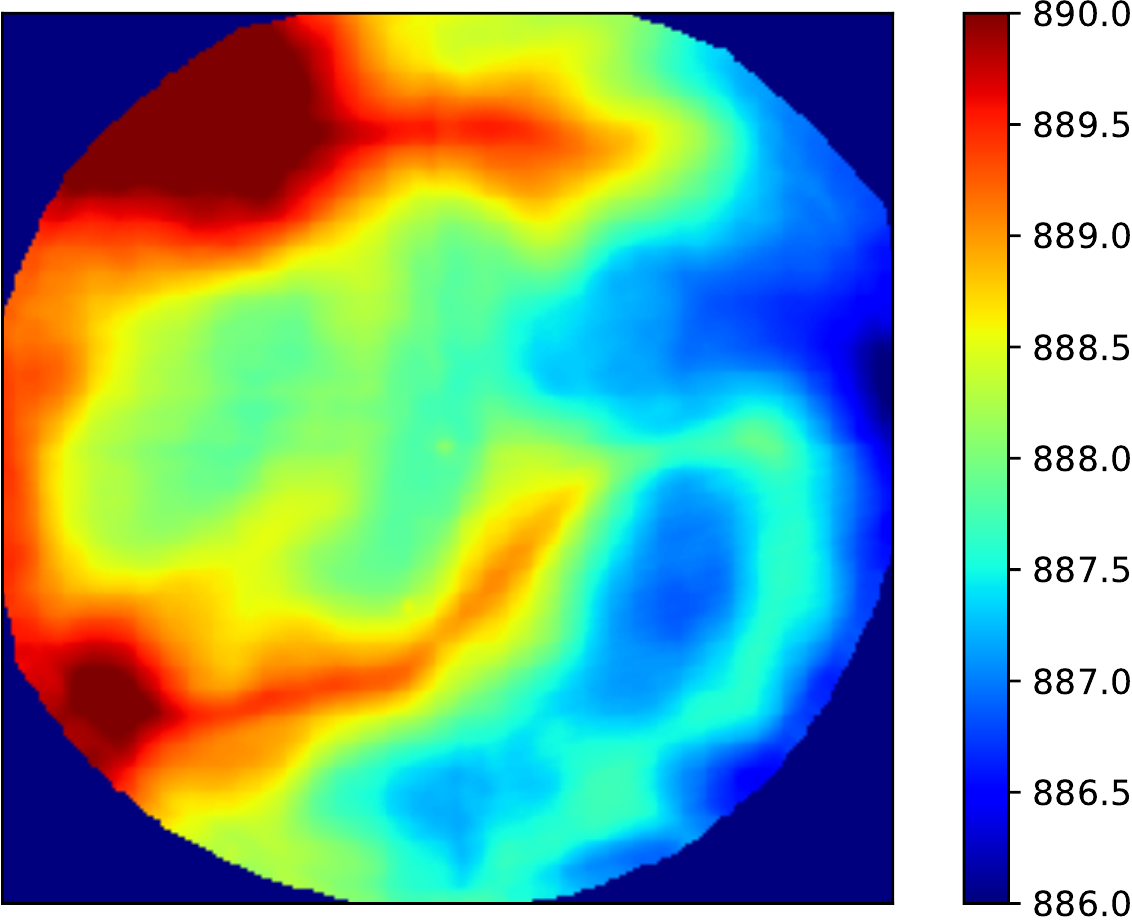}
        \caption{Elevation information of the field.}
        \label{fig:elevation_map_}
    \end{subfigure}
    \caption{Description of the study area.}
    % \label{fig:soil_hyd_pars} 
\end{figure}

The proposed irrigation scheduling approach was implemented to generate irrigation schedules for a large-scale agricultural field. The field in question, a Research Farm operated by Lethbridge College, spans an area of 26.4 hectares and has a circular shape. The center is located in Lethbridge, southern Alberta, with geographic coordinates approximately at latitude 49.72° N and longitude 112.80° W. A layout of the investigated field is shown in Figure~\ref{fig:layout_of_the_alberta_irrigation_center_}. The average elevation of the area is 888 meters. Regarding the soil composition, the predominant texture in the field is clayey loam, with occasional presence of sand lenses. To facilitate effective irrigation management, a five-span center pivot system has been installed in the field. This mechanized irrigation system follows a circular pattern around a central pivot and each span covers a distance of 294 meters. Furthermore, the center pivot system incorporates a commercial variable rate irrigation system, enabling variable irrigation application rates based on specific field requirements. To monitor soil moisture levels during the rotation cycle of the pivot, microwave radiometers have been installed on each of the five spans. These radiometers serve the purpose of accurately measuring the soil moisture content, providing valuable data for the irrigation scheduling process.

\begin{figure}[!t]
\centering
	\begin{subfigure}[b]{0.43\textwidth}
		\includegraphics[width=\textwidth]{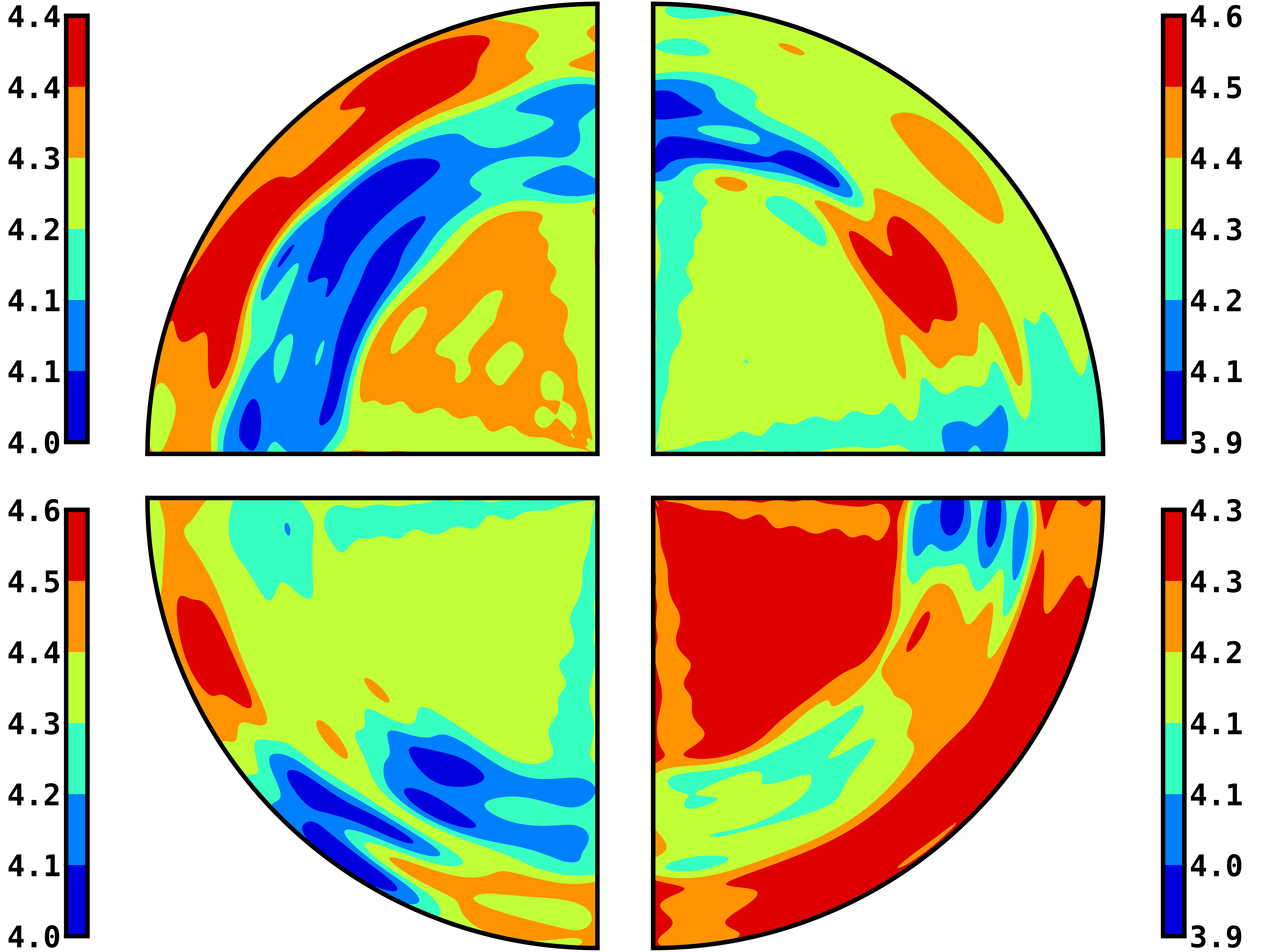}
		\caption{$\alpha~(m^{-1})$}
	\end{subfigure}
	\hspace{1.5cm}
	\begin{subfigure}[b]{0.43\textwidth}
		\includegraphics[width=\textwidth]{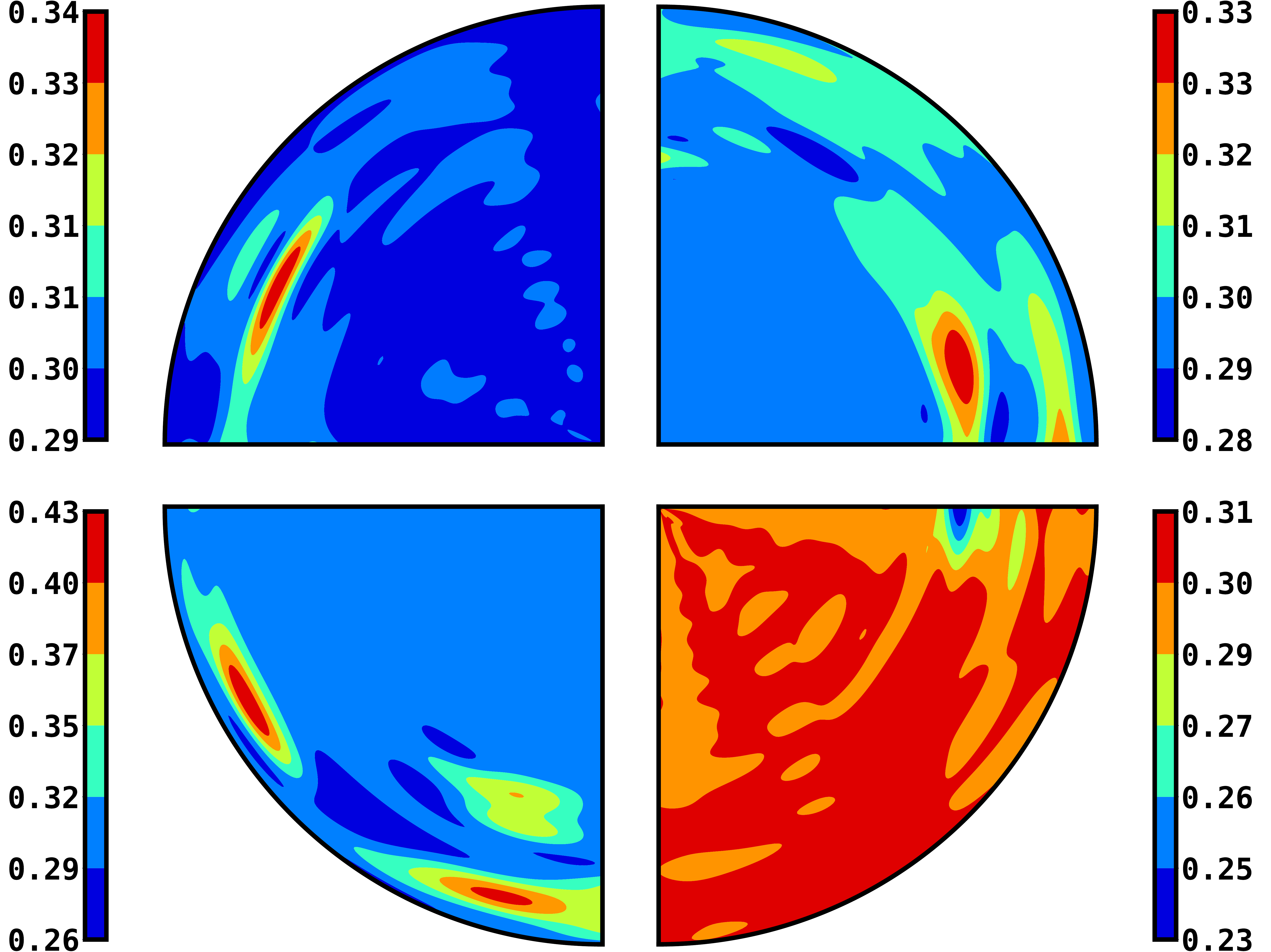}
		\caption{$K_s~(m \cdot day^{-1})$.}
	\end{subfigure}
	\caption{Spatial maps of some estimated hydraulic parameters.}
	\label{fig:soil_hyd_pars} 
\end{figure}

The surveyed region exhibits varying elevations, ranging from 886 meters at the lowest point to 890 meters at the highest point. An interpolated field elevation map, depicted in Figure~\ref{fig:elevation_map_}, visually represents this elevation distribution. It is therefore crucial to consider the elevation information when delineating the considered field into irrigation management zones to effectively capture water movement within the field.

To prepare for the implementation of the irrigation scheduling approach, soil moisture observations were collected using microwave remote sensors installed on the center pivot of the field. These observations were utilized to estimate both the soil water content and soil hydraulic parameters across the entire field. The estimation process covered the period from May 16, 2022, to August 31, 2022. By leveraging the available soil moisture data during this period, estimates of soil moisture and various soil parameters were derived for the field under study. For a concise overview of the estimation procedure, readers may refer to Section \ref{delination}, while a comprehensive description of the steps involved can be found in \cite{agyeman2022simultaneous}. In this section, we present two of the estimated parameters: the saturated hydraulic conductivity ($K_s$) and the curve fitting parameter ($\alpha$), to provide a brief overview of the estimation results while maintaining brevity. The spatial maps of these two estimated hydraulic parameters are depicted in Figure \ref{fig:soil_hyd_pars}. The figure illustrates the variability of soil hydraulic parameters across the field. It is important to note that the five estimated soil hydraulic parameters, along with elevation information of the study area, serve as the primary attributes for the delineation of the field into management zones.

\subsubsection{Management zone delineation of the study area}
\begin{figure}
  \centerline{\includegraphics[width=0.95\textwidth]{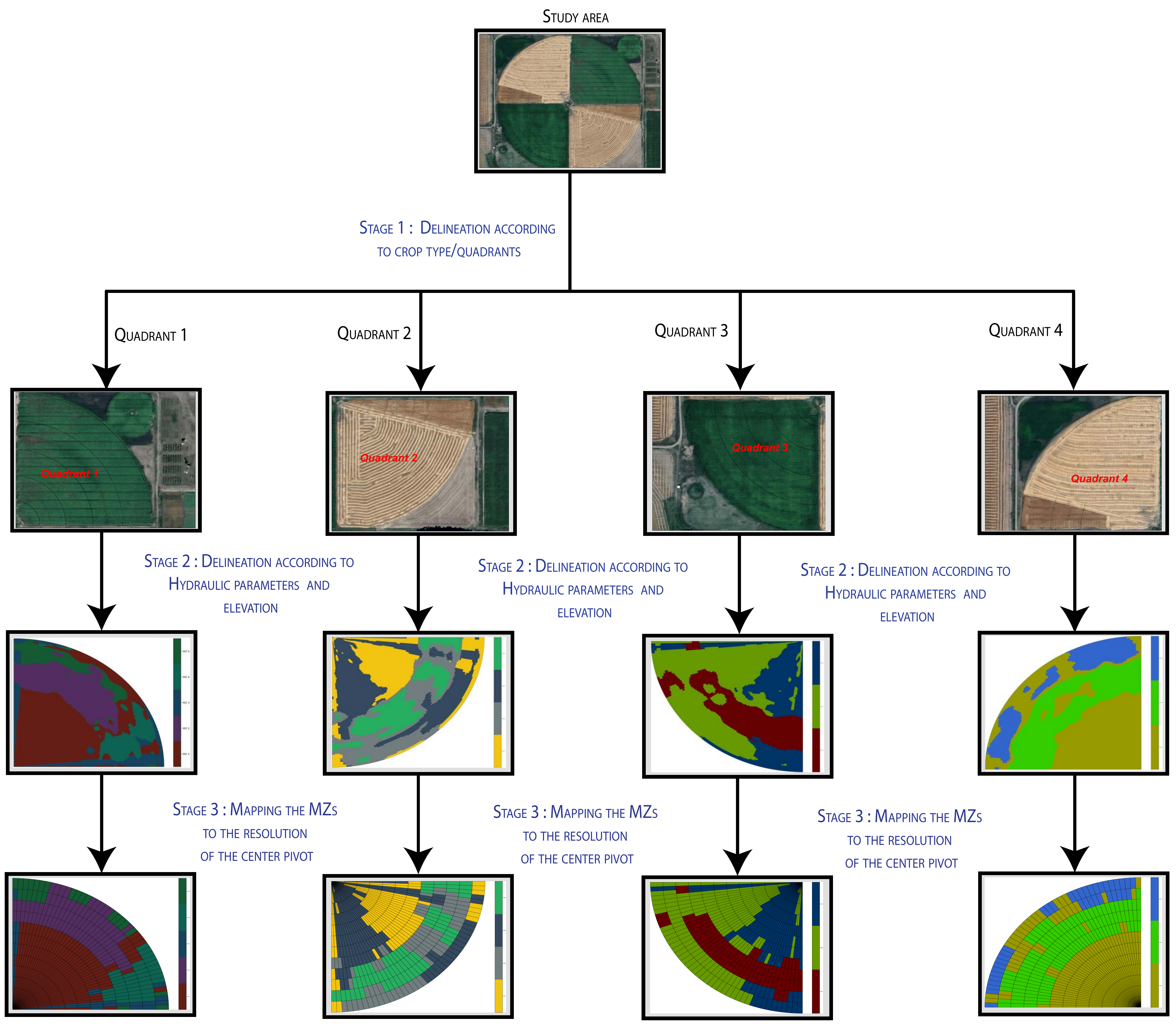}}
  \caption{\centering{Three-stage irrigation management zone delineation approach.}} 
  \label{fig:management_zone_delin}
\end{figure}

\begin{figure}
	\begin{subfigure}[b]{0.4\textwidth}
		\includegraphics[width=\textwidth]{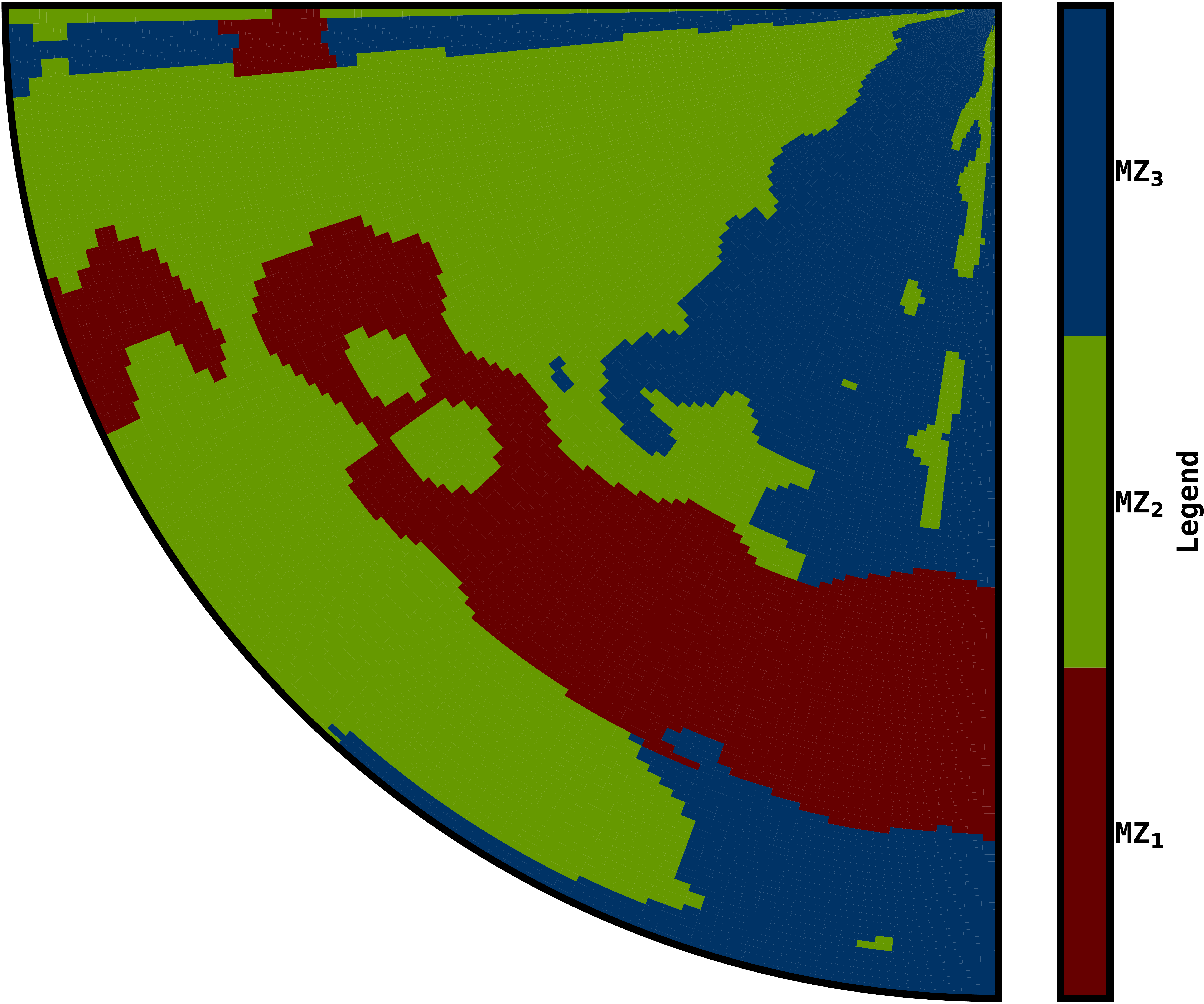}
		\caption{Irrigation management zones.}
	\end{subfigure}
	\hspace{3cm}
	\begin{subfigure}[b]{0.4\textwidth}
		\includegraphics[width=\textwidth]{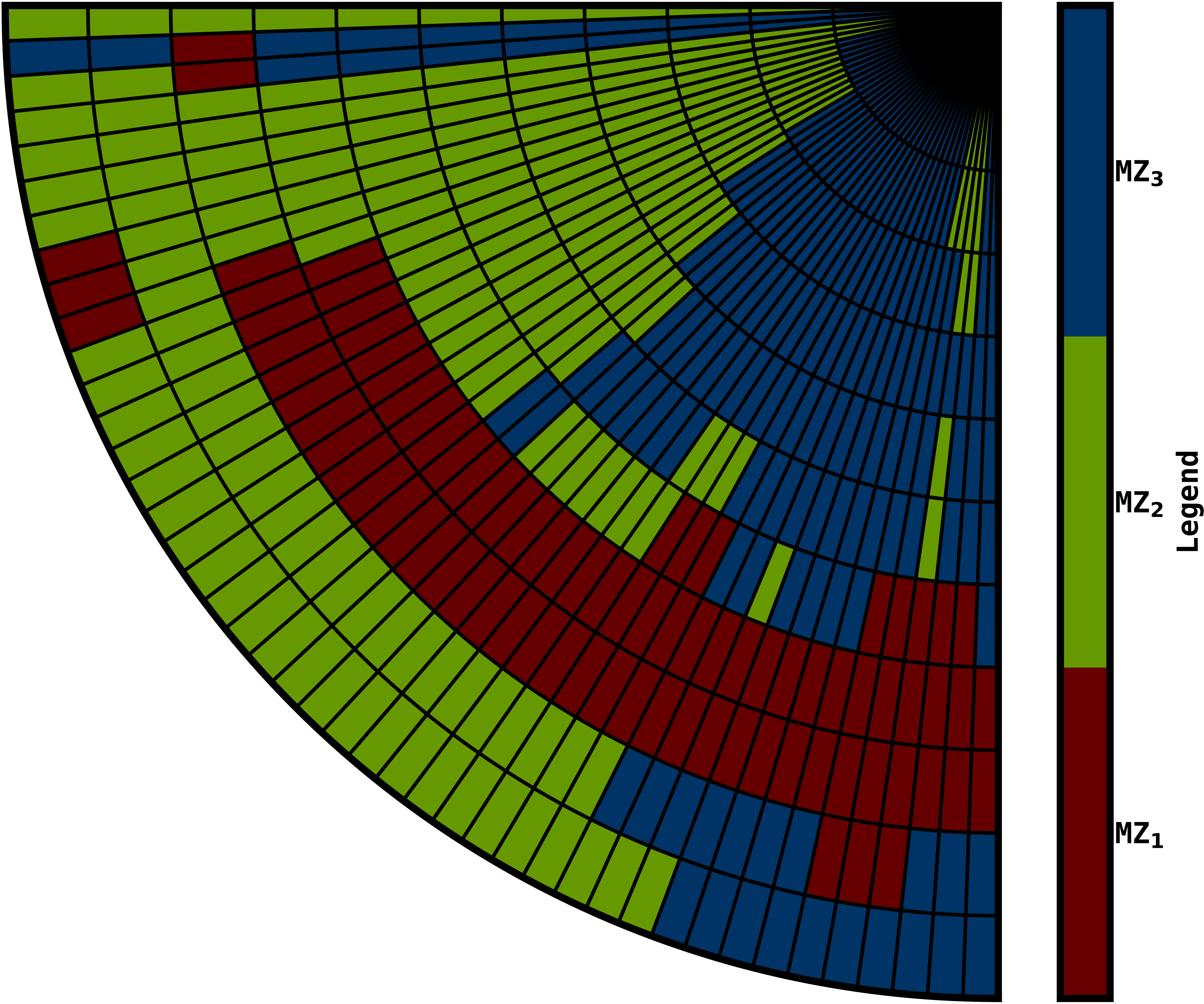}
		\caption{Pivot resolution representation.}
	\end{subfigure}
	\caption{Irrigation management zones and corresponding pivot resolution representation of the zones.}
	\label{fig:Q3_management zones}
\end{figure}

Typically, in the studied field, various types of crops are planted in the four quadrants. As a result, the initial division of the field based on crop type leads to four irrigation management zones, corresponding to the four quadrants of the circular field. After this initial division, each quadrant was further subdivided into irrigation management zones based on estimated hydraulic parameters and elevation data. The resulting management zones were then mapped to match the resolution of the field's center pivot. Specifically, for each quadrant, the center pivot has a resolution consisting of 12 zones in the radial direction and 45 zones in the azimuthal direction. The entire field was divided into irrigation management zones according to the proposed procedure, as illustrated in Figure~\ref{fig:management_zone_delin}.

For simplicity and comprehensive analysis, the results for the third quadrant of the field are presented, examined, and discussed in detail. Figure~\ref{fig:Q3_management zones} displays the management zones of this quadrant. Figure~\ref{fig:Q3_management zones} (a) depicts the management zone map obtained through the application of the $k$-means clustering approach to the normalized dataset. It is evident that the clustering approach resulted in three distinct management zones. Management Zone 1 (MZ$_1$) was found to have an average soil hydraulic parameter set corresponding to loamy soil, with an associated elevation of 889 m. Similarly, Management Zone 2 (MZ$_2$) exhibited an average soil hydraulic parameter set of loamy soil, but with an average elevation of 888 m. In Management Zone 3 (MZ$_3$), the average soil type was identified as sandy clay loam, and the average elevation measured 888.5 m. Figure~\ref{fig:Q3_management zones} (b) displays the spatial distribution of the three irrigation management zones, which were mapped to align with the resolution of the field's center pivot.

Subsequently, the remaining steps of the proposed scheduling approach are specifically implemented on the third quadrant of the field. This decision is made to ensure simplicity in the analysis process while maintaining the generalizability of the results. Focusing on a single quadrant allows for a thorough examination of the outcomes obtained through the proposed approach. It should be noted that applying the same steps to the remaining quadrants can be done easily and directly, without any loss of generality.

\subsection{Simulation settings}
In this section, we provide a comprehensive account of the application of the remaining steps of the proposed approach to the study area, which encompasses an actual agricultural field. Our specific focus lies on Quadrant 3, a representative section of the field, allowing for a detailed analysis while preserving the applicability of the results to the broader context. During the simulation processes, various parameter values and data were crucial in shaping the outcomes. We delve into the specific details of these simulation settings, which encompassed the development of LSTM models, the training of RL agents, and the subsequent evaluation of the scheduler. 
% The evaluation of the scheduler was performed over the course of two growing seasons, providing an extensive perspective on the performance and effectiveness of the approach in real-world agricultural scenarios. 
This comprehensive analysis of the simulation settings serves as a foundation for understanding the subsequent results and discussions, enabling a deeper understanding of the applicability and performance of the proposed approach in practical agricultural systems.

\subsubsection{LSTM model development}
\begin{table}[t]
	\caption{Simulation settings of the calibrated 1D Richards equation during the open-loop simulations.}
	%\small % Font size can be changed to match table content. Recommend 10 pt.
	\centering
	\begin{tabular}{cc}
		\toprule
		\textbf{Depth of soil}& 1.0 m\\
		\hline
		\textbf{Number of spatial nodes}& 31\\
		\hline
		\textbf{Spatial discretization} & \makecell{Top 0.50 m with 21 nodes \& 0.025 m spacing \\ Bottom 0.50 m with 11 nodes \& 0.05 m spacing}\\
		\hline
		\textbf{Time step size} & 30 minutes\\
		\hline
		\textbf{Temporal discretization}& Backward Differentiation Formula\\
		\hline
		\textbf{Process uncertainty}& \makecell{Normally distributed with \\Mean = 0.0 \& Standard deviation = 0.0005}\\
		\bottomrule
	\end{tabular} \label{tbl:sim_settings}
\end{table}

\begin{table}[t]
	\caption{Specification of the LSTM networks.}
	%\small % Font size can be changed to match table content. Recommend 10 pt.
	\centering
	\begin{tabular}{cc}
		\toprule
		\textbf{Number of hidden layers}& 2\\
		\textbf{Number of LSTM units in each hidden layer}& 400\\
		\textbf{Number of epochs}& 40\\
		\textbf{Learning rate} & 0.0001\\
		\textbf{Optimizer}& Adaptive moment estimation (Adam)\\
		\textbf{Sequence length}& 5\\
		\textbf{Loss function}& Mean squared error\\
		\bottomrule
	\end{tabular} \label{tbl:hyper_pars}
\end{table}
In order to simulate the soil moisture dynamics in the three management zones of Quadrant 3, the initial step involved applying the Richards equation to describe the soil moisture dynamics within each management zone. The calibration process of each Richards equation, for each management zone, entailed using the hydraulic parameters corresponding to the cluster's centroid. Subsequently, LSTM networks were developed for each management zone using a dataset obtained from  extensive open-loop simulations of the calibrated Richards equation. These simulations encompassed the generation of random initial states, reference evapotranspiration, crop coefficient, and irrigation rate inputs. The simulation settings utilized during the open-loop simulations of the Richards equation can be found in Table~\ref{tbl:sim_settings}. To ensure accurate irrigation scheduling for the target crop type, which in this case was spring soft wheat, the model training phase incorporated two different rooting depths: 0.5 m and 1.00 m. For all three management zones, the reference evapotranspiration and crop coefficient were randomly generated within the ranges of 0.1 mm/day to 8.99 mm/day and 0.4 to 1.02, respectively. The irrigation rate was randomly generated between 4.0 mm/day and 52.0 mm/day for MZ$_1$, 4.3 mm/day and 59.6 mm/day for MZ$_2$, and 5.0 mm/day and 62.3 mm/day for MZ$_3$. It is important to note that historical weather data and local irrigation management practices informed the selection of input ranges for the open-loop simulations of the Richards equation.
Following the resampling of the open-loop simulated data to a daily time frame and the computation of root zone soil moisture content using the methodology outlined in Section~\ref{lstm_mod_dev}, the LSTM models were trained using the Keras Library in Python. The model architecture and hyperparameters employed during the training process are provided in Table~\ref{tbl:hyper_pars}, with the note that these values were determined through experimentation.

\subsubsection{Training the RL agents}
\begin{table}[t]
    \caption{List of PPO hyperparameters and their values.}
    %\small % Font size can be changed to match table content. Recommend 10 pt.
    \centering
    \begin{tabular}{cc}
        \toprule
        \textbf{Hyperparameter}&  \textbf{Value}\\
        \midrule
        Horizon (T)& 30\\
        Learning rate & 0.0001\\
        Minibatch size  & 64\\
        Number of epochs  & 10\\
        Discount & 0.99\\
        Generalized advantage estimation parameter & 0.97\\
        Clipping parameter & 0.25\\
        Entropy coefficient & 0.01\\
        Total Number of Episodes & 300000\\
        \bottomrule
    \end{tabular} \label{tbl:hyper_pars_rl}
\end{table}
\begin{table}[t]

  \begin{center}
  \caption{Parameters of formulations $\mathbb{P}(y)$ and $\mathbb{P}_{\text{M}}(y)$ }
  \begin{tabular}{cc}
  \toprule
    \textbf{Parameter}& \textbf{Value} \\     \midrule
    $R_u$ & 9000\\    
    $R_c$ & 1000\\     
    $z_r$ (m) & 0.50 and 1.00 \\   
     {$\bar{Q}$ (MZ$_1$, MZ$_2$, MZ$_3$)} & 22000000, 22000000, 22000000\\  
     {$\underline{Q}$ (MZ$_1$, MZ$_2$, MZ$_3$)} & 20000000, 20000000, 20000000\\  
     {$\theta_{\text{fc}} / \bar{\nu}$ (MZ$_1$, MZ$_2$, MZ$_3$)} & 0.280, 0.280, 0.300\\
     {$\theta_{\text{pwp}}$ (MZ$_1$, MZ$_2$, MZ$_3$)} & 0.120, 0.120, 0.160\\    
     {$\underline{\nu}$ (MZ$_1$, MZ$_2$, MZ$_3$)} for MAD = 40\% & 0.216, 0.216, 0.244\\     
     {$\underline{\nu}$ (MZ$_1$, MZ$_2$, MZ$_3$)} for MAD = 65\% & 0.176, 0.176, 0.209\\     \bottomrule
     % $\hat{y}_0$ (MZ1, MZ2, MZ3) &  0.157, 0.145, 0.165\\     \hline
  \end{tabular} \label{tbl:parameter_values}
  \end{center}
\end{table}
The summary of the final parameter values for the formulations $\mathbb{P}(y)$ and $\mathbb{P}_{\text{M}}(y)$ used in training the RL agents across all three management zones in Quadrant 3 is provided in Table~\ref{tbl:parameter_values}. It is important to note that the per unit costs in these formulations were considered as tuning parameters in this study. Tuning these costs involved exploring multiple scenarios by adjusting the per unit costs in $\mathbb{P}(y)$ and $\mathbb{P}_{\text{M}}(y)$. Through an iterative process, the RL agents' performance was evaluated under different cost configurations, leading to the determination of the most suitable values for achieving the desired irrigation management objectives. The values of the volumetric moisture content at field capacity and permanent wilting point for the various soil types present in the management zones were obtained from Reference~\cite{huffman2012irrigation}.

To represent the policy of each agent, a fully connected multilayer perceptron with two hidden layers and a hyperbolic tangent (tanh) activation function was employed. The decision for irrigation action was modeled using a Categorical distribution, while the irrigation rate action was modeled using a Gaussian distribution. The hyperparameters utilized during the training of the agents, as well as other settings specific to the Proximal Policy Optimization (PPO) algorithm, are summarized in Table~\ref{tbl:hyper_pars_rl}. The training of the RL agents was conducted using the Tensorforce library in Python.

\subsubsection{Scheduler evaluation}
\begin{figure}[ht]
    \centerline{\includegraphics[width=0.70\columnwidth]{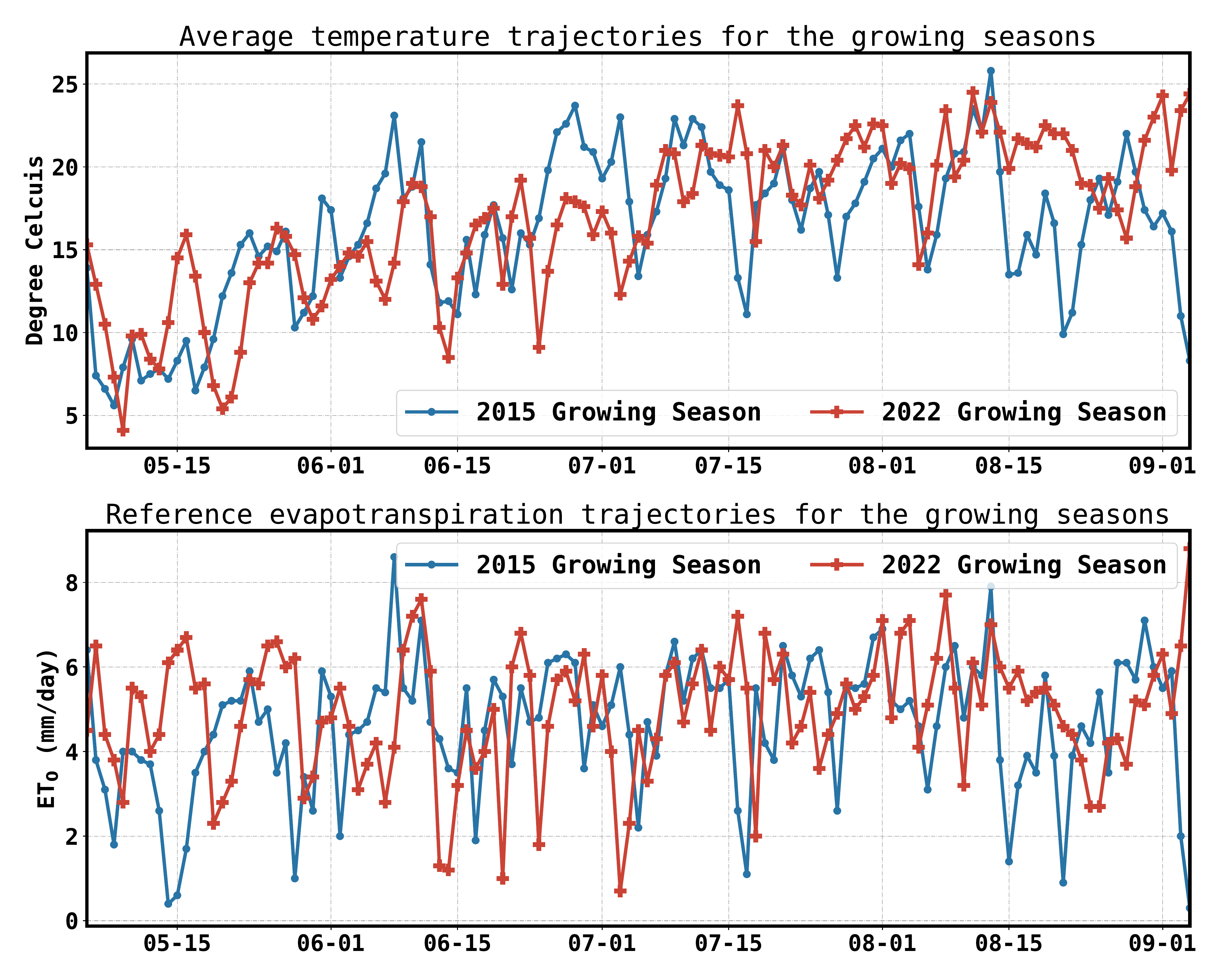}}
    \caption{Relevant weather data for the considered growing seasons.} 
    \label{fig:weather_data}
\end{figure}

The proposed scheduler was evaluated in the third quadrant over two growing seasons: the 2015 and 2022 seasons, considering moisture deficit MAD values of 40\% and 65\%. Figure~\ref{fig:weather_data} presents the relevant weather data utilized in this study, specifically from the selected growing seasons. It is important to note that a MAD value of 40\% is commonly used in the investigated field, but it may be considered conservative. Therefore, the scheduler was further assessed using a less conservative MAD value of 65\% to evaluate its ability to ensure optimal plant growth when allowing for more water depletion in the root zone. Additionally, it is worth mentioning that the 2015 season had lower total rainfall of 141.7 mm compared to the 2022 season which recorded a total value of 230.9 mm. These seasons were chosen to assess the scheduler's performance in relatively dry and wet conditions. Soft spring wheat was cultivated in the investigated quadrant during the 2022 season, and this crop was also considered for the 2015 growing season. To determine the crop coefficient ($\text{K}_{\text{c}}$) values for soft spring wheat during the simulation period, the following equation, calibrated for the study area, was utilized~\cite{bennett2011crop}:
\begin{equation}
\label{eq:kc_relation}
\text{K}_{\text{c}}(g) = -0.0207 + 0.00266g + \left(4.7\times10^{-8}\right)g^2 - \left(2.0\times10^{-9}\right)g^3 + \left(2.70\times10^{-13}\right)g^4
\end{equation} 
In Equation~\ref{eq:kc_relation}, $g$ is the cumulative growing-degree days (GDD). GDD is calculated as follows:
\begin{equation}
\text{GDD} = \text{T}_{\text{avg}} - \text{T}_{\text{base}}
\end{equation}
where $\text{T}_{\text{avg}}$ is the daily average/mean temperature and $\text{T}_{\text{base}}$ is the base temperature below which crop growth ceases (5\textdegree C). 

To determine the initial state for each management zone, the converged state estimates obtained from the estimation procedure outlined in Section~\ref{delination} were first mapped to the center pivot resolution of Quadrant 3. Within each management zone, the state estimates corresponding to the driest plot were selected as the basis for initializing the scheduler. It is expected that the initial soil moisture contents within a specific management zone would generally fall within a narrow range. Therefore, the irrigation rate calculated using the initial state of the driest plot was applied to the remaining plots within the same management zone. By adopting this approach, the scheduler starts with an appropriate irrigation strategy derived from the conditions of the driest plot and extends it to the other plots within the zone. 

After the initial state selection, the scheduler interacts with the calibrated 1D Richards equations of the three management zones, which serve as the representation of the real field during the evaluation. In the evaluation phase, a prediction horizon of 14 days is taken into account. While the true weather information, such as rainfall, reference evapotranspiration, and average temperature, is known during the simulations, uncertainty is incorporated into the weather forecasts used over the prediction horizon. This wad done to assess the robustness of the scheduler to uncertainties in weather forecasts. The uncertainty in the weather forecast is modeled as a normal distribution with a mean of 0 and a specified standard deviation. As the prediction horizon extends further into the future, the standard deviation values are gradually increased to reflect the increasing uncertainty associated with longer-term weather predictions. By incorporating this uncertainty, the scheduler's performance can be evaluated under realistic conditions, considering the inherent uncertainty in weather forecasting.  

Table~\ref{tbl:parameter_values} provides the relevant parameters of formulation $\mathbb{P}(y)$ used in evaluating the scheduler. The evaluation period for each growing season spanned from 5th May to 4th September, encompassing a total of 123 evaluations. In each evaluation, the three RL agents were initially assessed using relevant observations from all three management zones to obtain irrigation decision sequences for the prediction horizon. Subsequently, the binding irrigation sequence was determined, and three copies of formulation $\mathbb{P}$(y) were simultaneously evaluated to optimize the irrigation rates for each management zone using the interior point optimization solver. These optimized irrigation rates were then applied to the calibrated 1D Richards equation, and the resulting soil water information was used to initialize the subsequent evaluation of the scheduler. Throughout the evaluation process, a rooting depth of 0.5 m was considered from 5th May to 15th July. From the 16th of July to the end of the growing season, a rooting depth of 1.0 m was used to determine the irrigation schedules. It is worth noting that this practice is employed in the investigated field and was also adopted in this study. 

Both the triggered approach and the proposed approach were utilized to generate irrigation schedules for the selected growing seasons and specific MAD values. To enable a meaningful comparison between the two approaches, the total amount of water recommended by each approach was calculated for each growing season and MAD value. Additionally, the some relevant components of the cost function used in formulation $\mathbb{P}(y)$ were evaluated based on the results obtained from both scheduling methods. During the computation of these cost values, the parameter values presented in Table~\ref{tbl:parameter_values} were employed. Furthermore, the yield of spring soft wheat was predicted for each scheduling approach. For the evaluation of the yield, the parameters $Y_m$ and $k_y$ in Equation~\ref{eq:yield_eqn} were set to 8.8 Mg ha$^{-1}$ and 1.15, respectively. These values were obtained from Reference~\cite{bennett2011crop} and represent calibrated values specific to the study area. 

\section{Results and Discussion}
\subsection{Predictive performance of the identified LSTM models}
\begin{figure}[ht]
\centering
	\begin{subfigure}[b]{0.365\textwidth}
		\includegraphics[width=\textwidth]{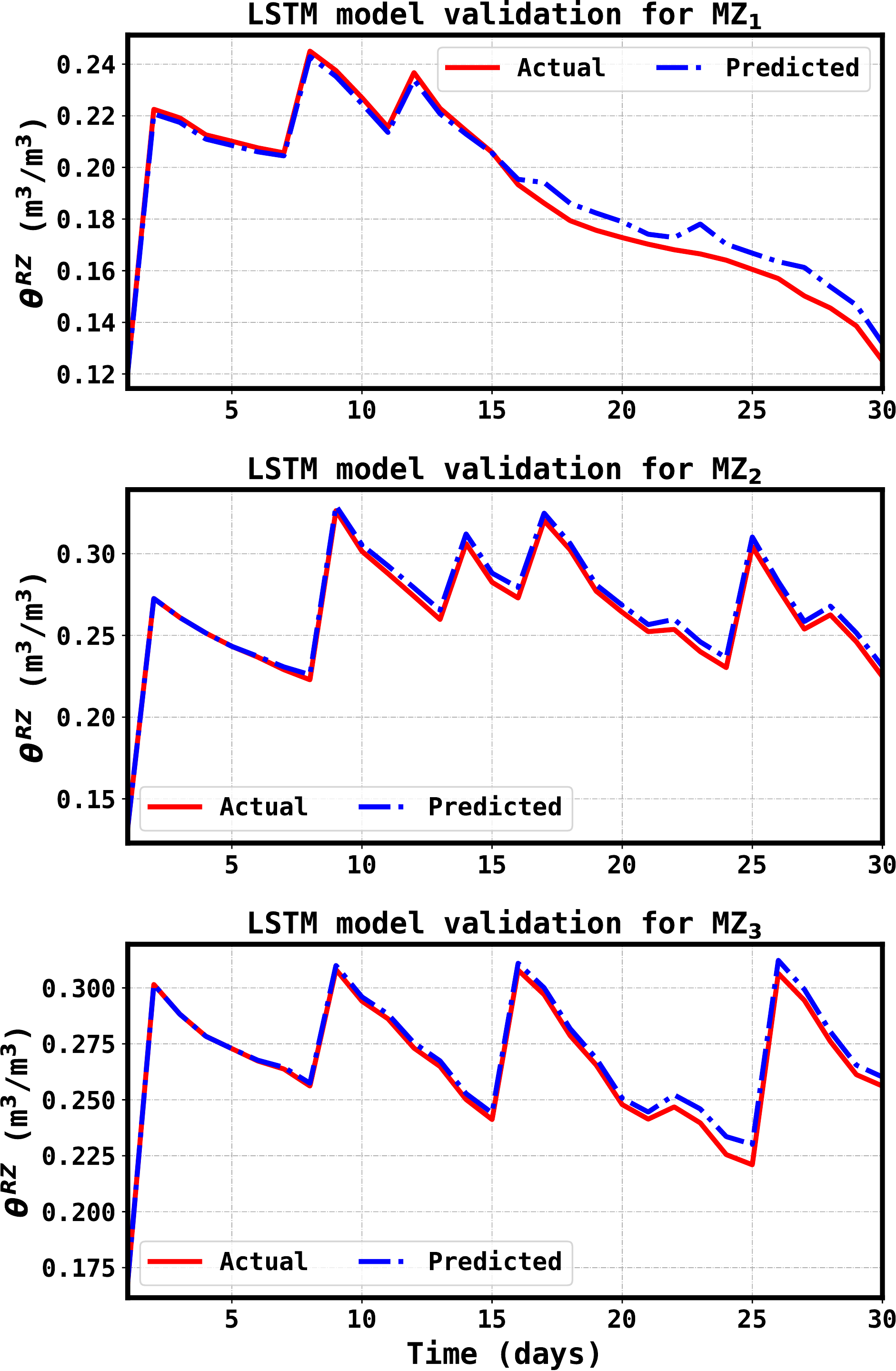}
		\caption{$\text{z}_{\text{r}}$ = 0.50 m.}
	\end{subfigure}
	\hspace{1cm}
	\begin{subfigure}[b]{0.36\textwidth}
		\includegraphics[width=\textwidth]{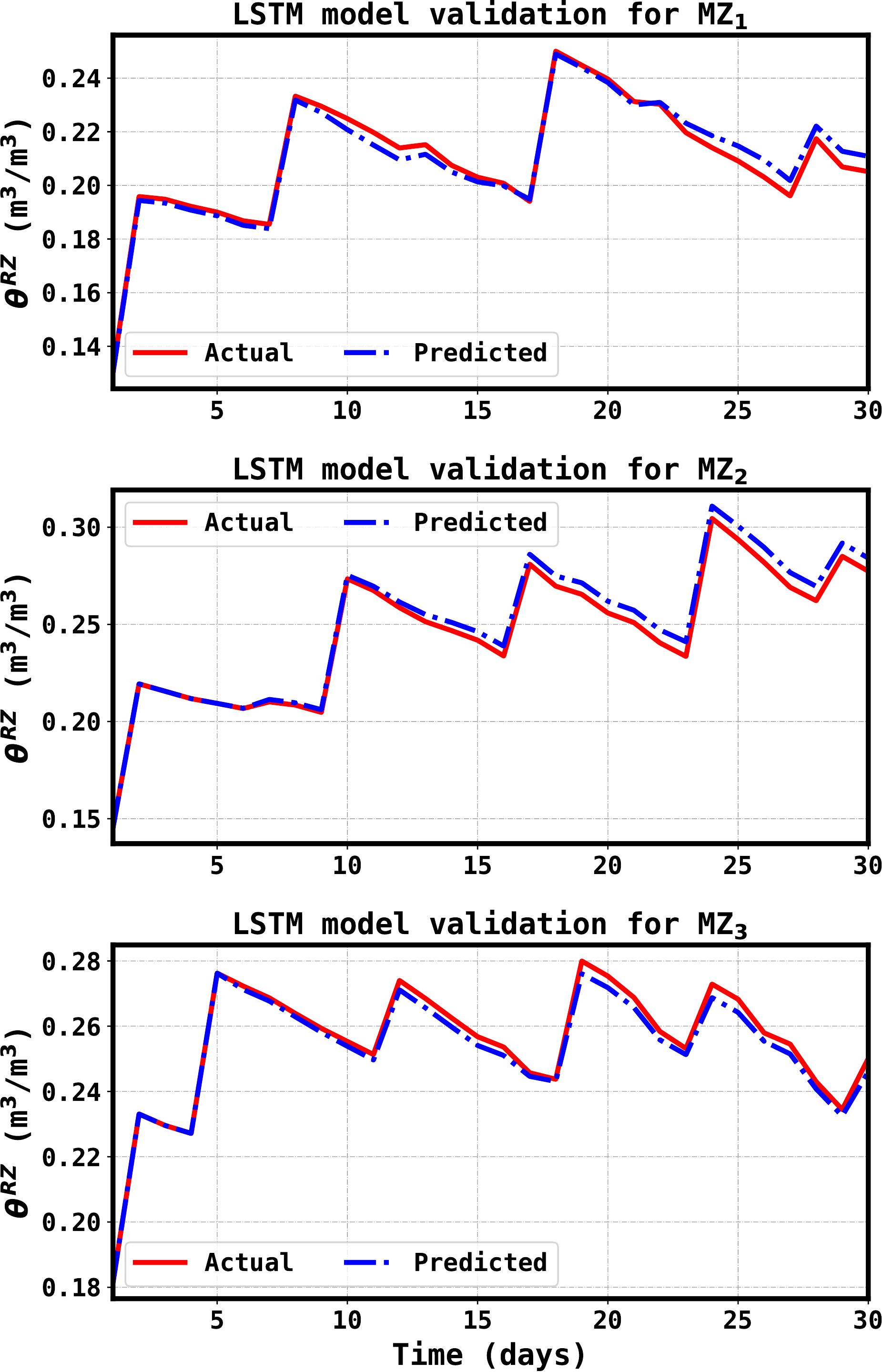}
		\caption{$\text{z}_{\text{r}}$ = 1.00 m.}
	\end{subfigure}
	\caption{Predictive performance of the identified LSTM models for the considered rooting depths.}
	\label{fig:model_perform} 
\end{figure}

To assess the performance of the LSTM models developed for each management zone and rooting depth, a comparative analysis was conducted against a simulation experiment using the calibrated Richards equation. The experiment spanned 30 days and involved selecting various initial states within each management zone, along with predefined trajectories for the irrigation rate, crop coefficient, and reference evapotranspiration inputs. These same inputs were applied to the LSTM models, and the resulting predictions were compared with the weighted average of the spatial volumetric moisture content values obtained from the Richards equation for each management zone. The results of this experiment, depicted in Figure~\ref{fig:model_perform}, demonstrate that the LSTM models successfully capture the trend in soil moisture dynamics as obtained from the Richards equation. This observation is reinforced by the high R$^2$ values exceeding 0.90 for all management zones and depths considered. Additionally, the RMSE values presented in Table~\ref{tbl:predictive_comp} confirm the accuracy of the identified LSTM models in predicting root zone soil moisture.

The results obtained from the experiment highlight the accuracy of the identified LSTM models in making multi-step-ahead predictions. In this experiment, the LSTM models, which were trained to predict one step into the future, were applied recursively over the investigated period. This recursive approach involved using the output at a specific time step as an input in the subsequent time step. Such a recursive framework is essential in model predictive control algorithms, as it enables the prediction of process outputs several time steps ahead at each time step.

The obtained results in this section provide evidence that the proposed LSTM model for the calibrated 1D Richards equation in each management zone is both robust and accurate. These findings align with several other studies that have also utilized LSTM models to describe soil moisture dynamics in agro-hydrological systems~\cite{agyeman2023lstm,adeyemi2018dynamic}. Moreover, the results suggest that the universal approximation capability of neural networks can be effectively leveraged to directly model root zone soil moisture content. This novel approach is expected to offer practicality and computational efficiency compared to previous studies, such as~\cite{capraro2008neural}, where separate neural network models were trained to estimate soil moisture content at various depths in the soil profile, and the root zone soil water content was subsequently determined by weighted averaging of the predicted values at each depth.
\begin{table}[t]
	\caption{Predictive performance of the identified LSTM models over a period of 30 days.}
	\centering
	\begin{tabular}{cccc}
		\toprule
		\textbf{Rooting depth} ($\text{m}$)& \textbf{Management zone description} & \textbf{RMSE} ($\text{m}^3/\text{m}^3$) & \textbf{R}$^2$ (-) \\
		\midrule
		0.50& 1 &  0.0053 & 0.975 \\
		0.50& 2 & 0.0044& 0.985\\
		0.50& 3 & 0.0037 & 0.984 \\
		1.00& 1 & 0.0033 & 0.978\\
		1.00& 2 & 0.0048 & 0.979 \\
		1.00& 3 & 0.0023 & 0.985\\
		\bottomrule
	\end{tabular} \label{tbl:predictive_comp}
\end{table}

\subsection{Evaluation of the RL Agents}
\begin{figure}[ht]
  \centerline{\includegraphics[width=0.65\columnwidth]{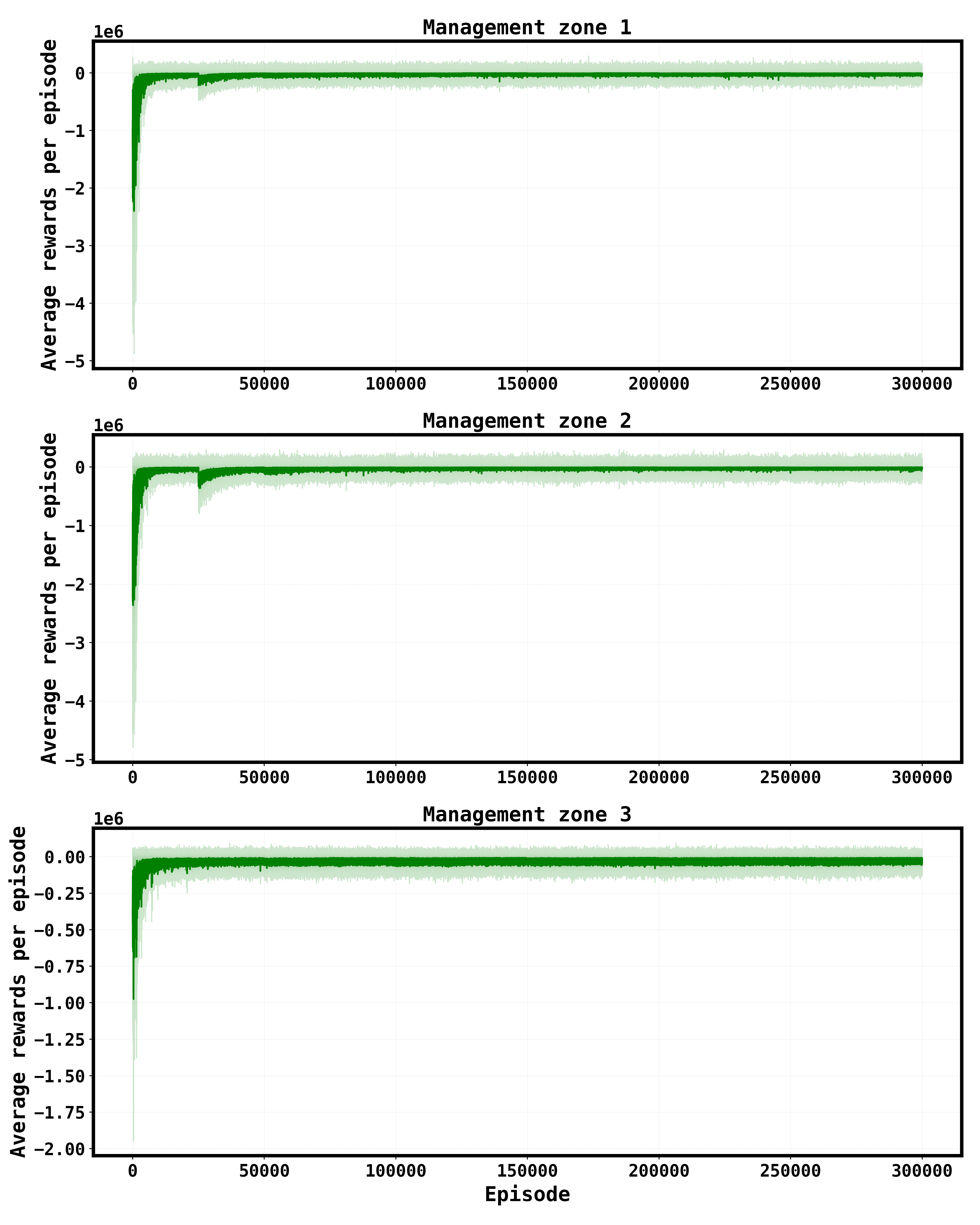}}
  \caption{Trajectory of the average episodic rewards for the RL agents for each management zone.} 
  \label{fig:rew_traj_rl}
\end{figure}

For each management zone, the RL agents were trained to generate irrigation schedules at two different moisture deficit levels (MADs): 40\% and 65\%. To provide a concise summary, the trajectories of the rewards obtained during the training stage for the MAD of 40\% are presented. Figure~\ref{fig:rew_traj_rl} demonstrates the average episodic rewards obtained after running each agent 10 times, showing an upward trend that indicates an improvement in agent performance over successive training iterations. Additionally, the RL agents exhibit convergence to stable and optimal policies, as evidenced by the rewards reaching a plateau during the training stage. Overall, the trained RL agents can effectively provide optimal irrigation decisions tailored to their respective management zones.

The utilization of the RL paradigm for determining binary variables in mixed-integer programs is a noteworthy addition to the  array of approximation methods, including the sigmoid approximation~\cite{de2005transmission} and time-step grouping~\cite{richards2005mixed}, that can be employed to simplify these programs and reduce their solution time. In the field of irrigation scheduling, sigmoid approximation has been employed to approximate binary variables in specific studies~\cite{agyeman2023lstm,agyeman2023IFAC}. However, despite its promise, sigmoid approximation introduces approximation error, compromises interpretability, necessitates careful parameter selection, can lead to non-convexity, and encounters challenges related to scalability and adaptability. On the other hand, the RL approach addresses these limitations by directly learning optimal policies for binary variables, eliminating approximation errors, enhancing interpretability through learned decision-making rules, adapting parameter values through exploration and exploitation, handling non-convex problems, scaling efficiently to large-scale scenarios, and adapting to dynamic environments. Although RL entails challenges such as training complexity, it offers a comprehensive solution to overcome the limitations associated with sigmoid approximation, establishing itself as an effective approach for handling binary variables in mixed-integer problems.

\subsection{Evaluation of the proposed irrigation scheduler}
\begin{figure}[ht]
	\centerline{\includegraphics[width=\columnwidth]{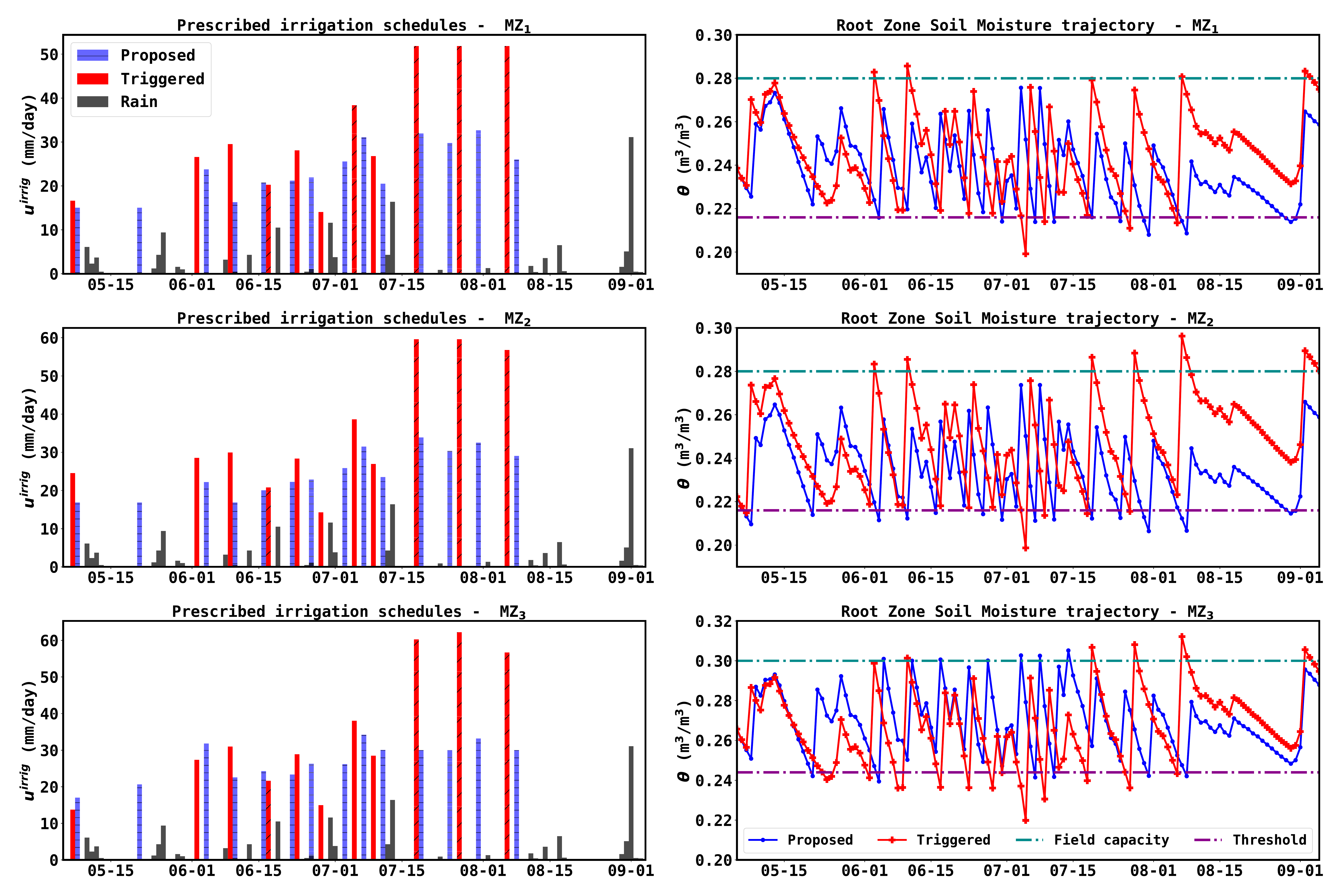}}
	\caption{Prescribed irrigation schedules and the trajectories of root zone soil moisture content under the schedules for the 2015 season (dry) with MAD=40\%.} 
	\label{fig:dry_mad_40}
\end{figure}

\begin{figure}[ht]
	\centerline{\includegraphics[width=\columnwidth]{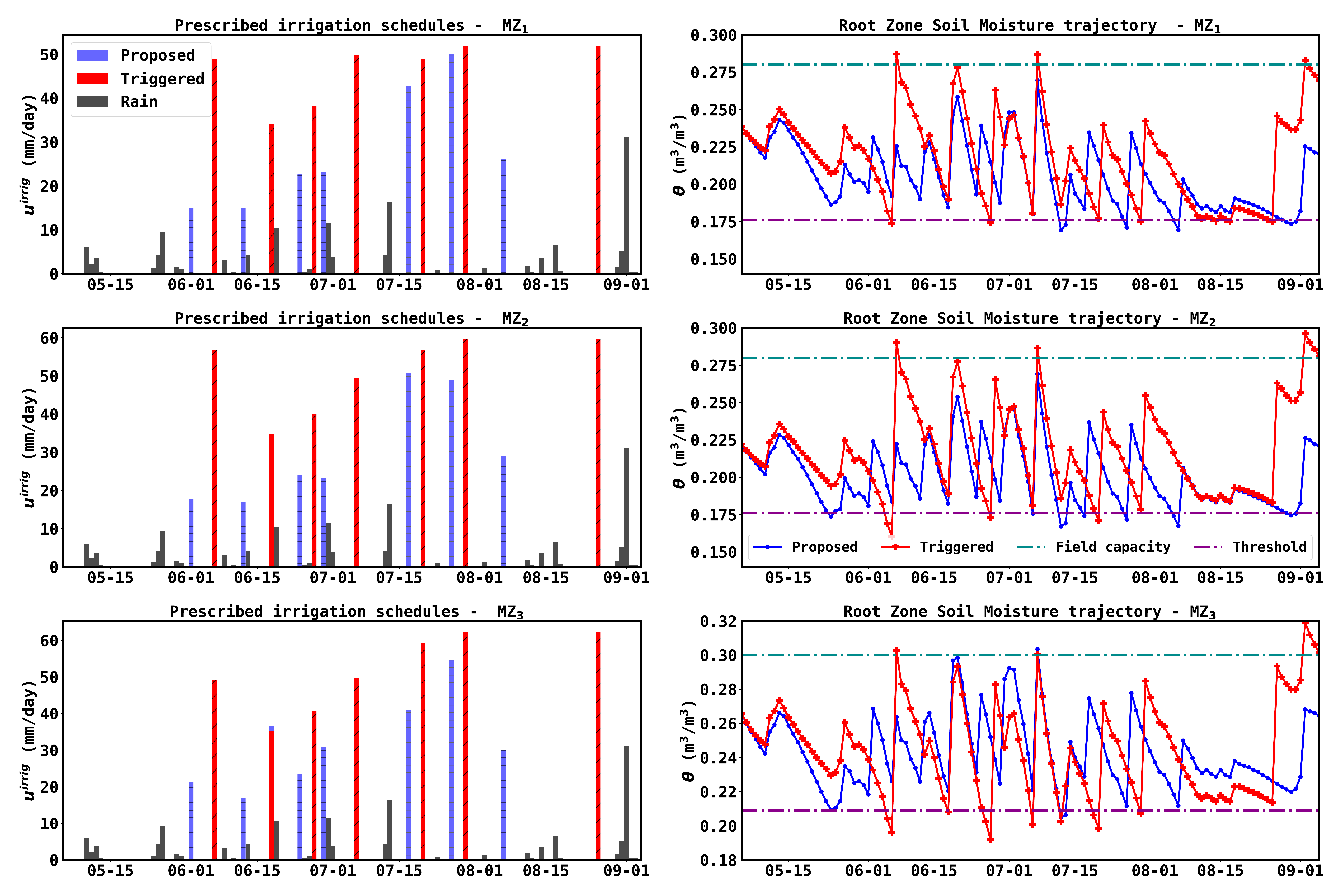}}
	\caption{Prescribed irrigation schedules and the trajectories of root zone soil moisture content under the schedules for the 2015 season (dry) with MAD=65\%.} 
	\label{fig:dry_mad_65}
\end{figure}

\begin{figure}[ht]
	\centerline{\includegraphics[width=0.85\columnwidth]{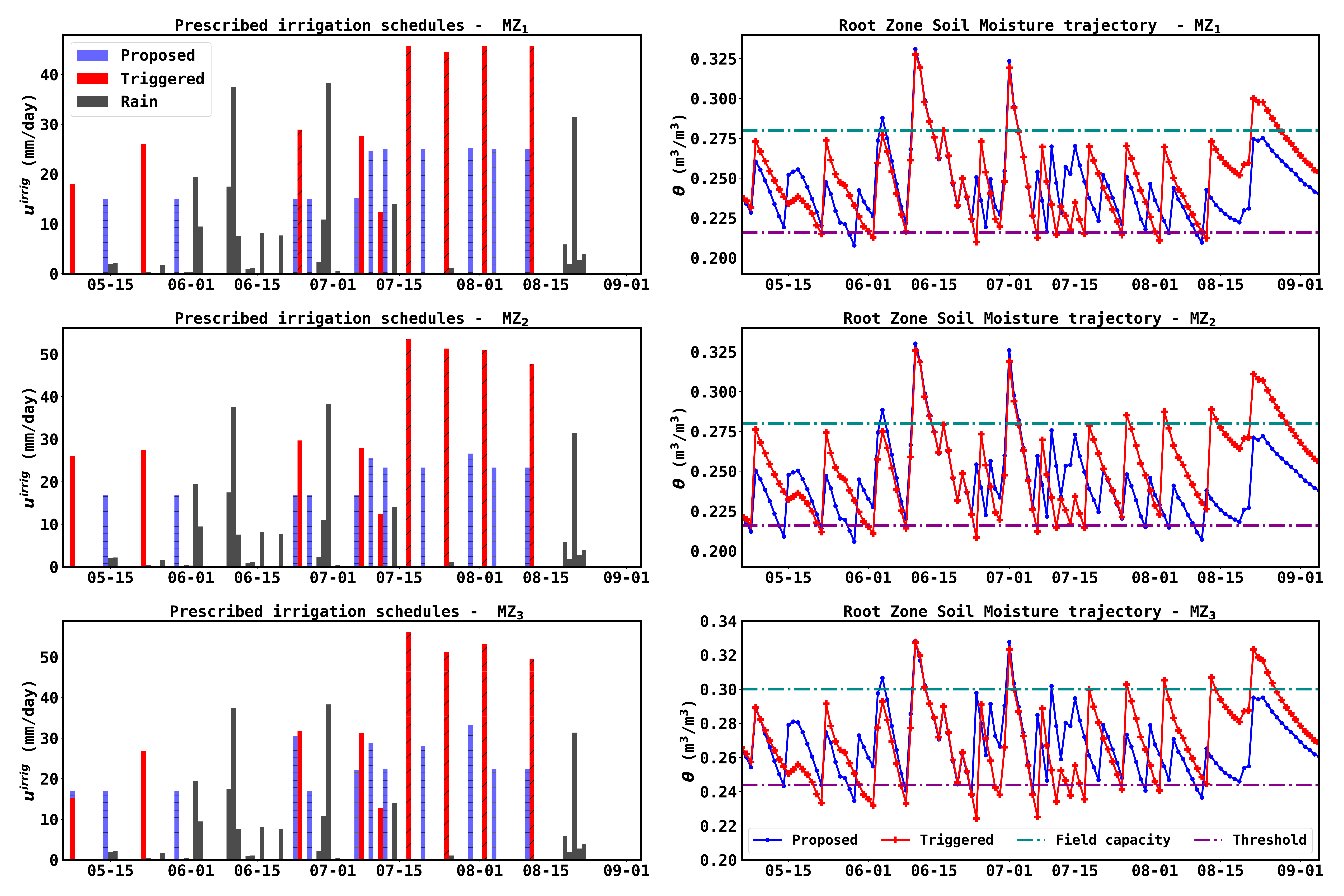}}
	\caption{Prescribed irrigation schedules and the trajectories of root zone soil moisture content under the schedules for the 2022 season (wet) with MAD=40\%.} 
	\label{fig:wet_mad_40}
\end{figure}

\begin{figure}[ht]
	\centerline{\includegraphics[width=0.85\columnwidth]{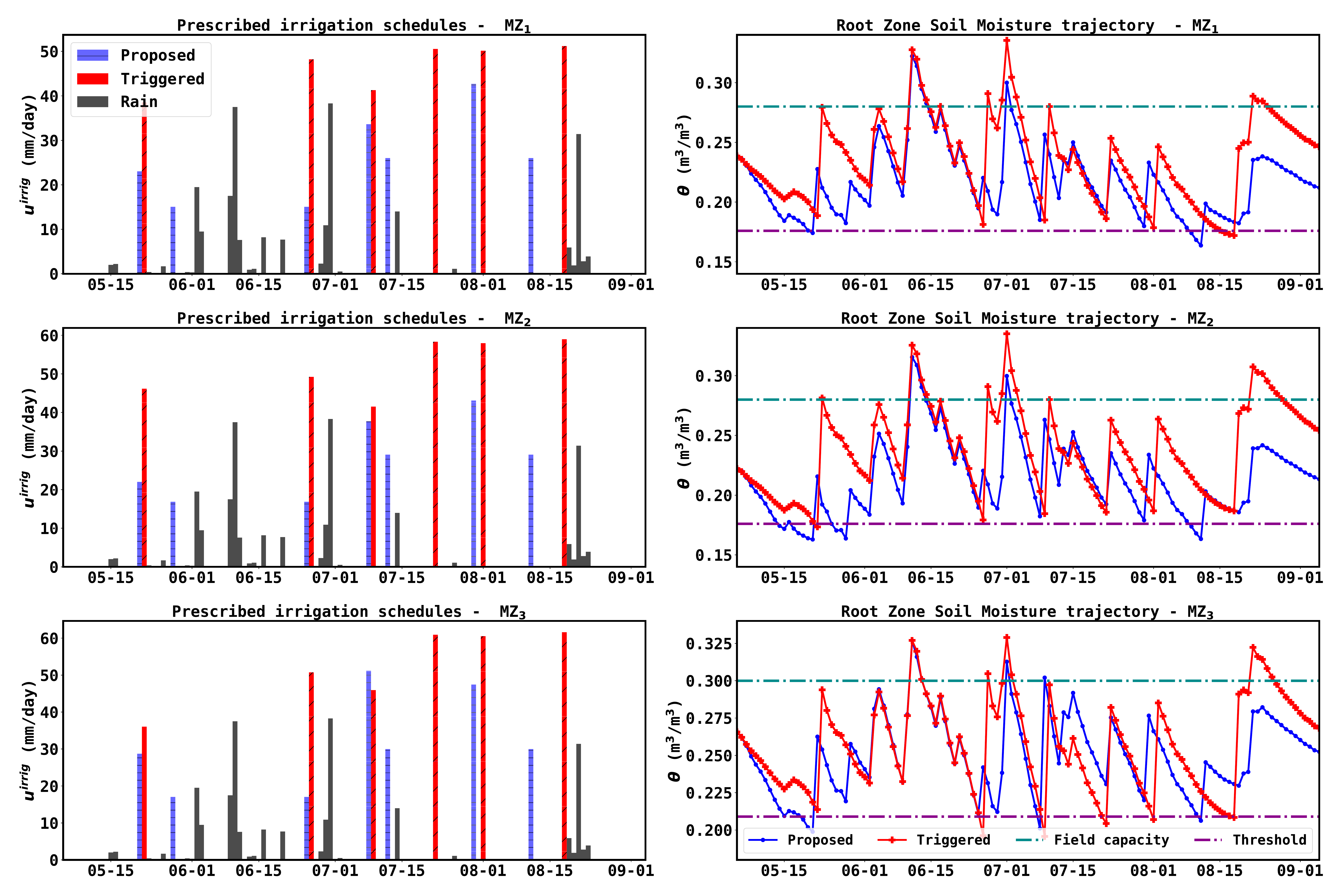}}
	\caption{Prescribed irrigation schedules and the trajectories of root zone soil moisture content under the schedules for the 2022 season (wet) with MAD=65\%.} 
	\label{fig:wet_mad_65}
\end{figure}

Figures ~\ref{fig:wet_mad_40} and ~\ref{fig:wet_mad_65} depict the prescribed schedules and the dynamics of the root zone soil moisture during the 2022 growing season for MAD (Moisture Availability Deficit) values of 40\% and 65\% respectively. Likewise, Figures ~\ref{fig:dry_mad_40} and ~\ref{fig:dry_mad_65} show the prescribed schedules and the corresponding root zone soil moisture dynamics for the 2015 growing season with MAD values of 40\% and 65\% respectively. Tables ~\ref{tbl:wet_40} and \ref{tbl:wet_65} present a summary of the total prescribed water and other relevant costs computed for the 2022 growing season, considering MAD values of 40\% and 65\% respectively. Similarly, Tables \ref{tbl:dry_40} and \ref{tbl:dry_65} provide the summary for the 2015 growing season with MAD values of 40\% and 65\% respectively.
As mentioned previously, the 2015 growing season experienced lower total rainfall compared to the 2022 season. Consequently, both scheduling approaches resulted in higher total irrigation rates for the 2015 season compared to the 2022 season. Additionally, the results demonstrated that for an MAD of 40\%, both scheduling approaches exhibited higher total irrigation rates compared to a less conservative MAD of 65\%. Similarly, both approaches made more frequent irrigation decisions for an MAD of 40\% compared to an MAD of 65\%. This observation aligns with the definition of MAD, as a lower MAD implies a smaller allowable depletion of root zone soil moisture, necessitating more frequent irrigations and higher total irrigation application depths.

Across all the scenarios considered, it is evident that the proposed approach achieved lower total irrigation rates compared to the triggered approach, while still ensuring higher predicted yields. The savings in total irrigation water achieved by the proposed approach were particularly pronounced when using a less conservative MAD value of 65\%. Furthermore, in all the scenarios, the proposed approach exhibited lower violations of the upper bound of the target zone compared to the triggered approach. This can be attributed to the proposed approach effectively utilizing rain forecasts and prescribing irrigation schedules that minimized violations of the upper bound. With the exception of the 2022 season long simulation for MAD=65\%, the proposed approach also recorded lower violations of the lower bound of the target zone compared to the triggered approach. The substantial violation of the lower bound in the triggered approach can be attributed to its simplistic subtraction of forecasted rain amount from the calculated irrigation rate without considering crop water consumption during the forecasted period, which is a myopic approach.

Furthermore, it is important to emphasize that the proposed approach achieves these advantages while employing more frequent irrigation events compared to the triggered approach. In the triggered approach, irrigation events are only initiated when the root zone soil moisture content falls below the lower bound of the target zone on a specific day. However, in the proposed approach, irrigation can be triggered on a particular day, even if the root zone soil moisture content is above the lower bound, based on the predicted behavior of the soil moisture over the prediction horizon. Although this leads to an increased number of irrigation cycles, the proposed approach takes a proactive stance by anticipating future violations of the lower bound and taking preventive action, resulting in improved irrigation management as evidenced by the results. Thus, despite prescribing more frequent irrigations, the proposed approach incurs lower overall costs in all the considered scenarios compared to the triggered approach.

The proposed approach also enables the use of less conservative MAD values without compromising crop yield. Employing a conservative MAD value of 40\% for irrigation scheduling serves to enhance the robustness of the scheduling approaches against disturbances that could potentially lead to crop stress. By using such conservative values, the threshold or lower bound of the target zone is set further away from the permanent wilting point, reducing the likelihood of reaching critical soil moisture levels. However, from the simulation results, it becomes apparent that less conservative MAD values can be employed, resulting in lower total irrigation rates and fewer irrigation cycles, while still maintaining crop health and minimizing the risk of reaching critical soil moisture levels. This suggests that the proposed approach offers flexibility in MAD selection, allowing for a more efficient utilization of water resources while ensuring optimal crop development.

	\begin{table}[t]
		\caption{Comparison between the triggered and the proposed approach for the 2022 season for MAD=40\%.  Arrows ($\downarrow$/$\uparrow$) denote percentage reductions/increases relative to the triggered approach.}
		\centering
		\begin{tabular}{cccc}
			\toprule
			\textbf{} & \textbf{Proposed}& \textbf{Triggered} & \textbf{Savings (\%)} \\
			\midrule
			 Prescribed irrigation (mm) & 884 & 949 & $\downarrow$ 6.8 \\     
		   Pivot rotations & 14 & 9 & $\uparrow$ 55.6 \\     
            Cost of violating $\theta_{\text{fc}}$ & 342840 & 423866  &$ \downarrow$ 23.6 \\ 
            Cost of violating $\theta_{\text{th}}$ &12872 &36086 & $\downarrow$ 66.2\\
            Overall cost &377672& 477493 & $\downarrow$ 26.4 \\
            Predicted yield (Mg/ha) & 8.33& 8.02 & $\uparrow$ 3.9\\ \bottomrule
		\end{tabular} \label{tbl:wet_40}
	\end{table}

	\begin{table}[t]
		\caption{Comparison between the triggered and the proposed approach for the 2022 season for MAD=65\%. Arrows ($\downarrow$/$\uparrow$) denote percentage reductions/increases relative to the triggered approach.}
		\centering
		\begin{tabular}{cccc}
			\toprule
			\textbf{} & \textbf{Proposed}& \textbf{Triggered} & \textbf{Savings (\%)} \\
			\midrule
			 Prescribed irrigation (mm) & 700 & 907 &$ \downarrow$ 22.8 \\     
		   Pivot rotations & 8 & 6 & $\uparrow$ 33.3 \\     
            Cost of violating $\theta_{\text{fc}}$ & 160100 & 460859  & $\downarrow$ 65.3 \\
            Cost of violating $\theta_{\text{th}}$ &27905 &7964 &$ \uparrow$ 250.4 \\ 
            Overall cost &202314& 482991 &  $\downarrow$ 58.1 \\ 
            Predicted yield (Mg/ha) & 8.40& 8.05 & $\uparrow$ 4.7 \\  \bottomrule
		\end{tabular} \label{tbl:wet_65}
	\end{table}
\begin{table}[t]
		\caption{Comparison between the triggered and the proposed approach for the 2015 season for MAD=40\%. Arrows ($\downarrow$/$\uparrow$) denote percentage reductions/increases relative to the triggered approach.}
		\centering
		\begin{tabular}{cccc}
			\toprule
			\textbf{} & \textbf{Proposed}& \textbf{Triggered Irrigation } & \textbf{Savings (\%)} \\
			\midrule
			 {Prescribed irrigation (mm)} & 1055 & 1127 & $ \downarrow$ 6.4\\   
		   {Pivot rotations} & 14 & 11  & $\uparrow$ 27.8 \\     
            Cost of violating $\theta_{\text{fc}}$ &  917 & 21115  &$ \downarrow$ 95.7 \\
            Cost of violating $\theta_{\text{th}}$ &11393 &36883 & $\downarrow$ 69.1 \\ 
            Overall cost &35807& 79137 & $\downarrow$ 54.7 \\ \
            Predicted yield (Mg/ha) & 8.49& 8.26 & $\uparrow$ 2.7\\ \bottomrule
		\end{tabular} \label{tbl:dry_40}
	\end{table}
	\begin{table}[t]
		\caption{Comparison between the triggered and the proposed approach for the 2015 season for MAD=65\%. Arrows ($\downarrow$/$\uparrow$) denote percentage reductions/increases relative to the triggered approach}
		\centering
		\begin{tabular}{cccc}
			\toprule
			\textbf{} & \textbf{Proposed}& \textbf{Triggered} & \textbf{Savings (\%)} \\
			\midrule
			 Prescribed irrigation (mm) & 885 & 1038 & $ \downarrow$ 14.7 \\   
		   Pivot rotations & 10 & 7 & $\uparrow$ 42.9 \\     
            Cost of violating $\theta_{\text{fc}}$ & 254 & 26521  & $\downarrow$  99.0 \\ 
            Cost of violating $\theta_{\text{th}}$ &7976 &22295 &$ \downarrow$ 64.2 \\ 
            Overall cost &26193& 65164 &  $\downarrow$ 59.8 \\ 
            Predicted yield (Mg/ha) & 8.73& 8.53 & $\uparrow$ 2.3 \\ \bottomrule
		\end{tabular} \label{tbl:dry_65}
	\end{table}


\section{Conclusion}
In conclusion, this study successfully demonstrated the effectiveness of integrating the three machine learning paradigms (unsupervised, supervised, and reinforcement learning) with model predictive control to design a precise irrigation scheduler for large-scale fields. Through the implementation of the k-means clustering approach in a three-stage process, irrigation management zones were precisely delineated to align with the resolution of the irrigation equipment, incorporating elevation and soil hydraulic parameters as attributes. Leveraging the universal approximation capability of neural networks, specifically long short-term memory models, this study directly modeled the dynamic behavior of root zone soil moisture content in each irrigation management zone. These models were trained using simulation data from calibrated 1D Richards equations, enabling accurate predictions of soil moisture dynamics. Based on the developed models, a mixed-integer model predictive control framework with zone objectives was designed to determine the daily irrigation decisions and rates, aiming to enhance soil water uptake while minimizing overall water consumption and associated irrigation costs. Furthermore, reinforcement learning was employed to identify optimal policies for the daily irrigation sequence over a predetermined time horizon. The introduction of a limiting management zone concept facilitated the derivation of a binding binary irrigation sequence, enabling simultaneous evaluation of the scheduler for all management zones. The application of the proposed approach to a large-scale field demonstrated its capability to achieve substantial water savings while ensuring crop health and development. The approach's flexibility also allowed for the utilization of less conservative MAD values, resulting in reduced irrigation frequency and improved water savings, while maintaining root zone soil moisture levels well above critical thresholds and guaranteeing optimal crop growth.

While the proposed approach shows promise, there are several modifications that can be made to further enhance its effectiveness. One priority is improving the optimality of the binding irrigation sequence by employing distributed techniques when training the reinforcement learning agents for each management zone. Utilizing a hierarchy of agents that communicate with each other can determine an irrigation decision sequence that is optimal for all management zones. Additionally, incorporating actual irrigation costs, including electricity costs, maintenance expenses, the actual cost of water used in irrigation, and the cost incurred when violating the bounds of the target zone, is crucial for more comprehensive decision-making. Furthermore, to enhance the overall robustness of the scheduler, it is advisable to implement it alongside a lower-level controller. This controller's task would be to track the prescriptions made by the proposed scheduler on an hourly basis. By integrating these components, the scheduler can adapt and respond in real-time to deviations or unforeseen circumstances, ensuring effective irrigation management. Implementing these modifications will contribute to the continued improvement and practical application of the scheduler, making it more effective and reliable in optimizing irrigation decisions while considering cost factors and maintaining system robustness.
\section{Acknowledgements}
Financial support from Natural Sciences and Engineering Research
Council of Canada and Alberta Innovates is gratefully acknowledged.

%\clearpage
%\bibliographystyle{ieeetr}
%\bibliography{references}
\end{document}